\documentclass[screen, acmsmall]{acmart}
\usepackage[inline]{trackchanges}

\usepackage{enumitem}

\newcommand{\jie}[1]{{\color{red}{#1}}}
\AtBeginDocument{%
  \providecommand\BibTeX{{%
    \normalfont B\kern-0.5em{\scshape i\kern-0.25em b}\kern-0.8em\TeX}}}

\setcopyright{acmcopyright}
\acmJournal{PACMHCI}
\acmYear{2023} \acmVolume{} \acmNumber{CSCW} \acmArticle{} \acmMonth{11} \acmPrice{15.00}\acmDOI{}

\begin{document}

\title[Twitch Users' Hate Raids Experience and Discussion]{Hate Raids on Twitch: Understanding Real-Time Human-Bot Coordinated Attacks in Live Streaming Communities}

\author{Jie Cai}
\affiliation{%
  \institution{Pennsylvania State University}
  \city{University Park}
  \country{USA}}
\email{jie.cai@psu.edu}
\orcid{0000-0002-0582-555X}

\author{Sagnik Chowdhury}
\affiliation{%
    \institution{New Jersey Institute of Technology}
  \city{Newark}
  \country{USA}}
\email{sc25@njit.edu}
\orcid{0009-0005-0674-5913}

\author{Hongyang Zhou}
\affiliation{%
    \institution{New York University}
  \city{New York}
  \country{USA}}
\email{hz2187@nyu.edu}
\orcid{0009-0005-2561-5401}

\author{Donghee Yvette Wohn}
\affiliation{%
  \institution{New Jersey Institute of Technology}
  \city{Newark}
  \country{USA}}
\email{yvettewohn@gmail.com}
\orcid{0000-0001-5583-4430}

\renewcommand{\shortauthors}{ Cai et al.}

\begin{abstract}
Online harassment and content moderation have been well-documented in online communities. However, new contexts and systems always bring new ways of harassment and need new moderation mechanisms. This study focuses on \textit{hate raids}, a form of group attack in real-time in live streaming communities. Through a qualitative analysis of hate raids discussion in the Twitch subreddit (r/Twitch), we found that (1) hate raids as a human-bot coordinated group attack leverages the live stream system to attack marginalized streamers and other potential groups with(out) breaking the rules, (2) marginalized streamers suffer compound harms with insufficient support from the platform, (3) moderation strategies are overwhelmingly technical, but streamers still struggle to balance moderation and participation considering their marginalization status and needs. We use affordances as a lens to explain how hate raids happens in live streaming systems and propose \textit{moderation-by-design} as a lens when developing new features or systems to mitigate the potential abuse of such designs.  
\end{abstract}





\maketitle
\jie{Preprint Accepted at CSCW 2023}
\section{Introduction}
Online abuse, also termed online harassment, is defined as ``pervasive or severe targeting of an individual or group online through harmful behavior'' \cite{DefiningTerms}. It is a prevalent and persistent problem for many online communities, from online forums decades ago to social media platforms like Facebook, Instagram, Reddit, and Twitter, to recent novel communities with real-time interaction like Twitch, Clubhouse, and Metaverse. It entails multiple harms to users \cite{Scheuerman2021AOnline} and those who deal with content \cite{Steiger2021TheSupport, Dosono2019ModerationCommunities}. Victims of harassment consider harassment an ongoing event and need various forms of support \cite{Goyal2022YouHarassment}.

In 2021, Twitch, a leading live streaming platform that provides multimodal interaction between broadcasters (streamers) and the audience (viewers), experienced a boycott by its users, primarily marginalized streamers, complaining about its inability to handle hate raids on its platform. Hate raids is a unique term developed for Twitch communities, originating from the ``Raids'' feature, which allows a streamer, after their stream, to send their viewers to other streamers' chatrooms to support each other's community. Hate raids occur when a streamer's channel is flooded with abusive/hateful messages from bot accounts \cite{Lee2022DontRaids}, or people use the raid mechanism to abuse a streamer \cite{Parrish2021HowRaid}. Several streamers started a ``\#DayOffTwitch'' campaign on Twitter to vent their dissatisfaction with the platform's slow and ineffective responses to these massive group attacks on marginalized streamers (e.g., black and LGBTQIA + streamers) \cite{Bennett2021WhatProtest}. Hundreds of bots joining the chatroom simultaneously disrupted the normal interaction among viewers with hateful message spam. Moreover, streamers and human moderators have to ban them with no end in sight.

While online harassment and content moderation are broadly investigated by HCI and CSCW scholars (e.g., \cite{Sultana2021Unmochon:Messenger,Chandrasekharan2017TheData,Wohn2017HowMedia, Chancellor2018NormsCommunities, Jhaver2019Human-machineAutomoderator,Kou2021PunishmentCommunity, Horsman2018APeriscope, McInnis2021ReportingWebsite}), the new affordances of a platform provide new forms of violations that have never been caught before and disrupt the interaction experience of users. Attackers are creative and always abuse the features of a system to cause trouble to the users on the platform, such as using the group voice chat on Discord to play porn to disrupt the voice discussion \cite{Jiang2019ModerationDiscord}. More broadly, as more marginalized and underrepresented groups also actively participate and diversify online communities, researchers have to be vigilant about exploring how the affordances of new technologies might be misused and abused \cite{Vitak2017IdentifyingHarassment}. 

In this study, we focus on coordinated group attacks in real time in live streaming communities, explore marginalized streamers' experience with \textit{hate raids} on Twitch, and identify the challenges the communities face with potential design implications to cope with these attacks. We ask:

\begin{itemize}
    \item RQ1: What are users' understandings and interpretations of hate raids?
    \item RQ2: What are the impacts of hate raids on live streaming communities?
    \item RQ3: What are the approaches and challenges to combat hate raids?
\end{itemize}

Through an analysis of scraped data on the Twitch subreddit (r/Twitch) about hate raids discussion, we contribute to understanding human-bot coordinated group attacks with  real-time nature and towards marginalized users in live streaming communities. We clarify that the targets of hate raids are mainly marginalized streamers, but the hate raids discussion in this study is from all affected groups (e.g., marginalized streamers, general streamers, viewers, moderators, and streamers' friends, and some tool developers). We use affordances as a lens to explain how attackers leverage live streaming systems to conduct hate raids and the harm and trade-off framework to explain marginalized streamers' sufferings. We also propose the \textit{moderation-by-design} concept (a concept suggesting that system design should always consider the potential abuse of such design and possible proactive and reactive responses to such abuse) to develop new features and moderation mechanisms and provide a list of recommendations and implications for stakeholders (platform, designers and developers, streamers, moderators, and viewers).

\section{Related Work}

\subsection{Harassment Towards Marginalized Groups and Content Moderation}
Many scholars in HCI have explored online harassment in different contexts, such as thread comments \cite{Goyal2022YouHarassment}, voice chat \cite{Jiang2019ModerationDiscord}, and social VR \cite{Blackwell2019HarassmentGovernance}. In all these contexts,  users experienced hate speech and harassment and suffered various intertwined harms: physical harm, such as self-injury and sexual abuse; emotional harm, such as depression and trauma; relational harm, such as damage to one's reputation and interpersonal relationships; and financial harm, such as loss of a digital asset or financial loss \cite{Scheuerman2021AOnline}.

Women and LGBTQ communities are often targets of online harassment \cite{Consalvo2012ConfrontingScholars,Keyton2007SexuallyConversation} and experience more harmful behaviors than men, such as physical threats and sexual harassment \cite{Brown2020LGBHarassment}, particularly when online communities are dominated by men and performance is perceived as masculine, as in gaming communities \cite{Braithwaite2014SeriouslyWomen}. Consequently, marginalized users are more likely to withdraw participation in online communities, or stay alone and anonymous with limited social signals to reveal their identities, such as using neutral avatars and avoiding using voice communication in gaming spaces \cite{Fox2017WomensStrategies,Kuznekoff2013CommunicationCues}. Marginalized users often lack of social support and experience emotional harms, such as anxiety and loneliness \cite{McLean2019FemaleStudy} and depression \cite{Lindsay2016ExperiencesAdults}. Their continued experience as "outsiders" urge cultural change in online environment \cite{cote2017can}.

Online harassment is either formed by an individual attacker or a group of attackers to initiate a hate campaign to attack a specific group, a marginalized group, or women in particular. Prior work does not clearly distinguish harassment by individual/random or group/organized. Organized harassment is less common than random attacks in many online communities and  different from random attacks in several aspects. First, scalability makes general moderation strategies impossible; there is no effective strategy to stop the attack, such as educating and communicating with individual attackers in general online harassment \cite{Cai2021ModerationCommunities}. Second, the intensity makes that no human moderator or algorithmic tool can effectively handle so much harmful content in a short time, and that the moderation action is less well planned and executed. For example,  the notable \textit{GamerGate} campaign on social media is a typical coordinated group attack on women in the video game industry in 2014 and 2015 \cite{Stuart2014ZoeLives}. Such large-scale online harassment is considered a semi-organized, pseudo-political movement on social networking sites \cite{Chatzakou2017HateTwitter}. The attackers are also less likely to be punished because it is challenging to detect their activities \cite{Chatzakou2017MeasuringBullying}, curb the mix of human and semi-automated bots to spread manipulative content \cite{Woolley2022DigitalInfluencers}, and promptly remove high-volume postings in communities \cite{Chatzakou2017HateTwitter}.

Content moderation refers to \textit{the governance mechanisms that structure participation in a community to facilitate cooperation and prevent abuse} \cite{Grimmelmann2015}. Many online communities more or less deploy content moderation that combines algorithmic tools and human moderators at the platform level; they also provide tools and features to end-users to customize the content (e.g., \cite{Jhaver2021EvaluatingTwitter}). While algorithmic tools are used to detect and react to harmful content at scale, they also augment human moderators' capability of monitoring content and reviewing specific instances \cite{Gillespie2020ContentScale, Lai2022Human-AIModeration}. Marginalized groups not only are easy to be targeted, but also experience disproportionate moderation \cite{Goyal2022IsAnnotation,Thach2022InvisibleReddit, Feuston2020ConformityModeration}. For example, trans and black social media users often experience account and content removal regarding their marginalized identities and are limited to present in the public sphere \cite{Haimson2021DisproportionateAreas}.

A thread of research has explored the moderation mechanism from an individual entity perspective, both the user's and moderator's views. From the end-user's perspective, many scholars have tested and prototyped tools to mitigate harmful content to end-users and their communities, such as the creator-led comment-filter tool on YouTube \cite{Jhaver2022DesigningModeration}, the language toxicity prediction and recommendation tool on Reddit \cite{Wright2021Recast:Visualization}, and evidence-capture tool on Facebook \cite{Sultana2021Unmochon:Messenger}. From the human moderator's perspective, a group of scholars has explored how moderators apply Twitch moderation tools to profile violators and manage viewers \cite{Cai2021AfterCommunities}, how moderators configure and collaborate with Reddit Automod \cite{Jhaver2019Human-machineAutomoderator}, and how interactive blurring tools mitigate moderators' exposure to harmful content \cite{Das2020FastContent}.

Another thread has explored the early detection mechanism at scale (e.g., \cite{Stringhini2015Evilcohort:Services,Nilizadeh2017POISED:Paths}). For example, on Reddit, researchers have developed a predictive model with graphs, users, community, and text features, to create an early warning system for human moderators to prevent inter-community aggressive behaviors \cite{Kumar2018CommunityWeb}; on YouTube, researchers have analyzed targeted videos' attributes to propose proactive moderation system to monitor hate attacks and mitigate their impact on content creators \cite{Mariconti2019YouAttacks}. In this study, we aim to understand how live streaming users, particular marginalized streamers, apply and evaluate the effectiveness of various moderation tools to deal with large-scale real-time coordinated group attacks.


\subsection{Technological Affordances and Online Harassment}

We use Norman's definition of affordances in HCI. Norman notes that affordances refer to the action possibility perceived by actors that an artifact offers for the action \cite{Norman1988TheThings}. Technology can have distinct impacts on users based on their interpretation, intention, and knowledge \cite{Hutchby2001TechnologiesAffordances}. In other words, users' online practices are not determined by technology, but by how they use it \cite{Norman1988TheThings}. An affordance exists once users perceive a feature/function and the potential actions associated with it \cite{Meredith2017AnalysingAnalysis}. The same technology can be used differently among users \cite{Jones2006ADecade, Page2022PerceivingMedia}. Thus, affordances can be shaped by both designers and users.

The Internet plays an essential role in shifting criminal opportunities from physical to virtual spaces \cite{Miro-Llinares2019WhatDrop}. Social networking sites can `afford' the attack because they provide communication channels between victims and attackers and allow attackers to collect and disseminate information about victims \cite{Mitchell2010UseUtilization.}. Social media has several specific affordances regarding content distribution: persistence (easily recorded and archived), replicability (easily copied), scalability (easily shared), and searchability (easily accessed by others) \cite{Boyd2008TakenPublics}. 

Vitak et al. \cite{Vitak2017IdentifyingHarassment} summarize two types of affordances that social media platforms may afford online harassment. The increased content visibility makes harassment reach broad audiences and encourages potential harassers to join harassment activities. Anonymity and pseudonymity also encourage users to act more hostile since their real identities are hidden online, and they fear less loss of reputation \cite{Friedman2001ThePseudonyms, Seering2017ShapingExample-Setting}. Therefore, under anonymity/pseudonymity, people would feel less responsible for their actions \cite{Sheng2020FromTwitch}. For example, research has shown that users who choose to be anonymous are more likely to post hateful comments; when an activity receives substantial hateful comments, it continues to receive such comments for a long time \cite{Zannettou2020MeasuringWebsites}. Marginalized users consider that anonymity and pseudonymity afford the safety for their community but also afford targeting and  abusing from outside attackers \cite{Scheuerman2018SafePeople}.

Attackers use affordances of platforms to innovate ways to behave negatively \cite{Robey2013InformationOdyssey}. For example, users often develop their understanding of the moderation system \cite{Jhaver2019DidReddit} and trick the algorithm with linguistic variations to avoid detection \cite{Kim2021TrkicAlgorithms,Chancellor2016thyghgapp:Communities}. Live streaming has some unique affordances, such as authenticity and synchronicity \cite{Cai2021ModerationCommunities} and is initially designed to share and engage with demographically distant users to form communities. However, it is also used to facilitate sexual abuse in children, as attackers use it to broadcast sexual content and distribute it globally \cite{Quayle2016ResearchingVulnerabilities, Horsman2018APeriscope}. Live chatroom for interaction also facilitates online harassment for the streamer as they stream with real-self and in real-time \cite{Uttarapong2021HarassmentNegativity}. In this study, we focus on hate raids on Twitch, a form of synchronous communication in the chatroom with hateful messages flow. We use affordances as a lens to explain how hate raids abuse the affordances of live streaming systems to harm marginalized users.

\subsection{Harassment and Content Moderation on Twitch} 

Live streaming as a novel media affords mass communication in the chatroom \cite{Hamilton2014StreamingMedia}. While streamers are broadcasting with various low- and high-fidelity equipment \cite{Drosos2022TheOpportunities}, viewers can simply register a pseudonymous account and send messages in the chat, and the streamer can read and respond orally to these messages. Oftentimes, the streamer would like to interact with the viewers to build communities \cite{Sheng2020FromTwitch} and promote prosocial behaviors in the chat \cite{Seering2019DesigningBehaviors}. Sometimes, attackers break the rules and start harassing the streamer/streaming content with toxic and hateful messages, even spamming these messages or emotes \cite{Riddick2022AffectiveLivestream}. Twitch, a leading live streaming platform, is perceived as a masculine space dominated by white and male streamers \cite{Consalvo2012ConfrontingScholars}. Marginalized streamers (e.g., women and LGBTQ+) suffer various online harassment, such as sexually lewd comments and hate speech related to racism, sexism, homophobia, and transphobia; they have to manage their emotions and apply human moderators and tools to deal with harassment \cite{Uttarapong2021HarassmentNegativity}. Hate raids as real-time coordinated group attacks exacerbate the aforementioned harassment with scalability and intensity by exposing the marginalized streamers in front of the camera.  Limited resources and tools to handle such situations in time constraints can potentially intensify the harms experienced by marginalized streamers as they watch all these happening in real time.

Twitch applies a multi-level moderation system to combat harmful content on its platform. At the platform level, Twitch not only works hard with AI development to detect harmful content, but also hires employees to actively monitor all streaming activities. As Twitch's VP of trust and safety said, “We combine proactive detection, and a robust user reporting system with urgent escalation flows led by skilled human specialists to address incidents swiftly and accurately” \cite{Grayson2022HowFacebook}. At the community level, it allows each streamer to appoint human moderators and apply moderation tools, such as Twitch AutoMod and third-party tools \cite{Cai2019CategorizingTwitch}, to manage the audience \cite{Wohn2020AudienceTwitch} and facilitate content moderation based on channel rules developed by the moderation team \cite{Cai2022CoordinationCommunities}. At the individual level, it provides settings for the viewer to filter the chat and block other viewers in the chat \cite{2021H2Report}.  

In this study, we focus on community-level moderation centered on streamers. Although a thread of research has explored moderation strategies to deal with harmful content \cite{Cai2021ModerationCommunities}, these strategies focus more on individual instances. Little is known about the generality of these strategies with regard to large-scale group attacks. We explore Twitch stakeholders' strategies and challenges to deal with hate raids.

\section{Methods}

\subsection{Data Collection}
In this study, we  collect comments on the Twitch subreddit (r/Twitch). Reddit is a large  public online forum and divided into "subreddits", which are communities focused on specific topics and allowing users to join and post thread and leave comments.  Posts and comments can be upvoted or downvoted, and can also receive awards. r/Twitch is the largest subreddit dedicated to Twitch, with approximately 1.2 million users at the time of data collection. This subreddit provides a neutral location for Twitch streamers, moderators, and viewers to share their experiences and seek advice on live streaming, such as how to set up live streaming equipment, what are the strategies to grow the viewership, how to deal with harassment and trolls in chatrooms, how to manage the viewership, and what are the updates about tools and policies from Twitch. We consider r/Twitch as the main source of data collection because (1) the first author has joined and followed the subreddit for more than two years; (2) hate raids mainly happened on Twitch, and r/Twitch is the initial and suitable place for Twitch users to discuss; (3) all timely discussion is archived with rich data types and samples, such as screenshot of hate raids on Twitch, resource to handle hate raids (e.g., external websites, shared files), streamers' complaint on other social media platforms.   

We used the R 3.0.5 package "RedditExtractoR" to search for threads by keyword.  We first use the terms ``hate raids'' and ``follow bot'' based on our observation of subreddit and news reports \cite{Lee2022DontRaids} and ``hate attack'' and ``mob'' in the literature review \cite{Mariconti2019YouAttacks}. We read threads output based on each keyword. In this phase, we also tested and confirmed that the package could capture all variations with only a single term (e.g., hate raid, hate raided, hate raiding, and HATE RAIDS generated the same output). Next,  we added new keywords based the output of first round reading, such as ``massive bot'' and ``group attack.''  To iteratively read and compare outputs of these keywords, We also removed some keywords with their variations such as ``repeated message'', ``raid'', and ``mob'' because these threads are about general spam and Twitch raid feature or irrelevant after reading the threads, which do not capture the nature of hate raids.  Hate raids as a new term unique to Twitch can cover most of the discussion if they are in the thread title and content.  Finally, we used the keywords: ``hate raid, follow bot, massive bot, hate attack, group attack.'' Then we merged the results, removed duplicates, and separated the results into two spreadsheets (Threads, Comments). The Threads contained the threads (the main posts), and the Comments contained the comments (replies to the main posts). The Threads spreadsheet includes the title (thread title) and text (detailed description of the topic). The Comments spreadsheet includes the text of the comment. We started data collection in January and collected 182 threads and 5392 comments in total till February 17th, 2022.


\subsection{Data Analysis}

We followed Fereday and Muir-Cochrane's six steps (codebook development, reliability test, initial theme identification, additional coding, theme identification, theme corroboration) and used a hybrid approach with inductive and deductive coding for the theme development \cite{Fereday2006DemonstratingDevelopment}.
First, three authors open-coded every thread on individual documents so that no author could influence another. If the threads (and later on the comments) contained a link to a different site, we followed the links and included their contents in the coding. Then, we shared the codes and discussed the similarities and differences between each rater's codes. By reconciling these codes, we generated a codebook with 20 meaningful codes plus two functional codes. Details are given in \autoref{codebook}.  

Particularly, we coded the threads ``not relevant (0)'' if they are not clearly related to hate raids. We removed those irrelevant threads and the comments related to these threads consequently. We also added a code called ``relevant but not in the list (22)'' to apply to the Comments spreadsheet coding, just in case we missed something in the codebook development process.  We finally kept 55 threads and 3,944 comments as final data for further analysis. Among the 55 threads, the earliest explicit description of hate raids was on 4/13/2021, but discussion exploded after a Public Service Announcement posted by r/Twitch moderators on 8/28/2021. Approximately 3\% of the posts we analyzed were from before 4/13 (they were about related issues but not hate raids themselves), 3\% were between 4/13 and 8/28/2021, and 94\% were after 8/28.  58\% of threads are created by streamers, 6\% by moderators, 10\% by viewers, and 26\% by unknown. The most commented thread had 465 comments.


Second, two coders independently applied the codebook to a random sample of 100 comments in the Comments spreadsheet. We used this to calculate the inter-rater reliability with Cohen's Kappa. At first, the Kappa was .75, showing moderately significant agreement between the authors. However, the two authors met to discuss this inconsistency and realized that it was because some comments had multiple codes. After deciding to use only one code per comment, the section was recoded, and the reliability was recalculated. The Cohen's Kappa was .90, an almost perfect agreement. Third, we initially developed themes based on the codebook to primarily familiarize the topics. Fourth, two coders independently coded the rest of the Comments spreadsheet. Fifth, after coding, the three authors worked together to organize the codes into subcategories and high-level themes with the inclusion of code (22). Lastly, we iteratively grouped  the categories into research questions with high quality examples for each category and adjust their fit.

\section{Results}
Hate raids have been regarded as a prominent experience of marginalized (i.e., people of color, LGBTQ, female, and disabled) Twitch streamers. Many streamers use racial/identity tags to connect with their communities and promote themselves. However, the tags make the marginalized group prone to being attacked since the hate raiders \textit{``target streams based on tags that were pro-LGBT or pro-equality.''} This finding supplements prior research about the Twitch tag that the tag system increase the identity-based visibility but may introduce new ways for LGBTQIA+ streamers to be targeted \cite{Lopez2022ToStreaming}. In this section, if we explicitly know the comments from a affected group,  such as streamer, viewer, and moderator, we mention it. If it is not clear, we use users in general. We use marginalized streamers and small streamers interchangeably to align with the literature and user discourse.  

Hate raids take place not only in the live chatroom but also off the stream because viewers can participate in chat even if the stream is not live: \emph{``these hate raids have been going to any offline channel do a bunch of stuff then report to Twitch that there's no modding chat.''} Hate raids can also migrate to other platforms after the live streaming on Twitch. Discord is a social media platform with voice and video calls and text messaging. Users can form a community called ``server'' with a collection of categories and channels for users to join and interact. Since many streamers have group chats in Discord to have off-stream interaction with their followers, mass follow bots join the Discord servers of the steamers and post disturbing images (e.g. \emph{``images of animal gore''}) and hateful words (e.g. \emph{``pinging @everyone with a message containing targeted harassment \& slurs''}).



\subsection{RQ1: Twitch Users' Understandings and Interpretations of Hate Raids}



\subsubsection{Mass Bot Follows}
Hate raids can be understood as mass bot follows. As a streamer experienced and summarized, \emph{``If you're online, it clogs up your followers' alerts, which could last for minutes or hours until you pause the alerts or hide the source in your software. If done when you are offline, it basically means that any followers-only mode would be easily bypassed.''} 

\begin{figure}
\includegraphics[width= 0.35\linewidth]{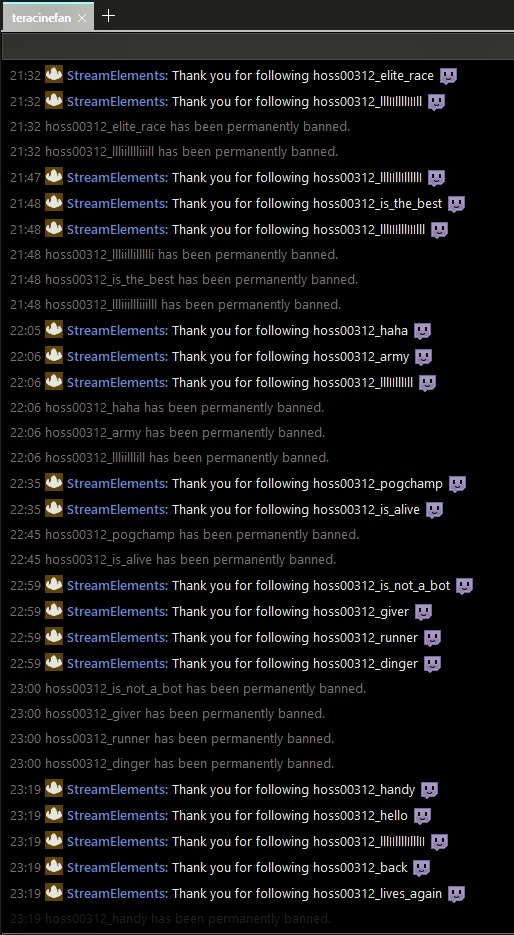}
\caption{A screenshot shared by a streamer at r/Twitch: the StreamElements (a moderation tool) notified new follows in the chatroom (in white color) and also banned hoss bots with notifications (in gray color).  However, bots came back with new names, such as hoss00312\_back and hoss00312\_lives\_again. Although there was no chat, the follow and ban notifications filled the chatroom.}
\label{fig}
\end{figure}

Many other small streamers shared their experiences with the ``hoss/host\_XXXX'' follow bots. These bots followed the streamers without posting or showing in the chat. As a small streamer said, \emph{``None of those actually do anything to combat. They aren't in chat/chatting, just following and unfollowing,''} as shown in \autoref{fig}.










Some follow bots intentionally triggered the alert notification by “following and unfollowing,” quoted from another streamer, \textit{``they are constantly causing the following alert animation and sound to kick off, degrading the stream quality and making my viewers stop wanting to watch and stopping me wanting to stream.''} By causing the notification sound, the ``hoss'' follow bots annoy the streamers and disrupt the streaming and viewing experience. Although there was a list of all the known over 1800 ``hoss'' follow bots shared among the streamers for their convenience to block the bots, some streamers argued that it was not helpful because \emph{``the bots rename themselves often enough that by the time you've banned all known bots, there are new ones.''}

\subsubsection{Mass Hate Messages by Follow Bots or/and Human}
Mass hate messages flooding the live chat in a short time can overwhelm the stream and ruin the community atmosphere. The messages can be sent by mass follow bots or humans. A user who believed that mass hate messages were caused by follow bots gave their definition of hate raids, \textit{``a hate raid is when an account sets up a stream with several bots viewing and posting the same message over and over again raids another channel where the bots continue to post that same message over and over again. The message is usually something inflammatory and insulting.''} 

Since streamers have encountered different cases of hate raids, they had different understandings and definitions of hate raids. While some streamers received mass hate messages \emph{``either shortly after (after a few seconds) or immediately after the bot has followed,''} some streamers commented that not all follow bots would cause hate raids. Some users suggested that mass hate messages could be a mix of bots and human behaviors. A user defined hate raids as \emph{``a mix between both raids and actual nasty people, and they target black people and people of the LBGTQIA+ community and basically spam a bunch of slurs and the n word.''}   

There have been controversies and discussions on the “raids” features on Twitch. A user described hate raids as \emph{“hundreds of bots suddenly following a streamer, and then posting heinous shit in chat. It doesn't have to be a Raid in Twitch language.”} There are some streamers with toxic community culture used the “raids” feature to pour their viewers into small streamers’ channels and subvert their community sphere. Counterintuitively, while the literal explanation of hate raids is “a raid with hateful messages”, hate raids do not have to involve the “raids” feature. According to a comment, \emph{“hate raids are raids usually containing bots that spam repeat messages in chat in large batches and groups. This does not use the actual raids feature to be performed.”}

\subsubsection{Human-engaged Streamer Entrapping}
Some hate raiders attacked streamers as a group. They entrapped the streamers, induced them to violate Twitch’s terms of services, and collected evidence to report the streamers to get them banned by Twitch. A viewer described a stream they watched that got a group attack: 
\begin{quote}

1. Someone created a new account on her [the streamer’s] Discord and posted disturbing/ graphic/ NSFW images. The images on the Discord screen showed up on the Twitch stream.

2. Someone on the Twitch stream said, `Those images are against the Twitch terms of services. Since you streamed those images on Twitch, I have no choice but to report you.' 

3. Someone created a new Twitch account meant to look like the streamer, but substituted a capital ‘i’ for a lowercase ‘L’, and then subscribed to her account, so it looked like ‘AlissaSmith just subscribed to ‘’AlissaSmith’ (example name, not the actual streamer). In the Twitch font, the capitol ‘I’ is almost indistinguishable from a lowercase ‘L’.

4. Someone on another Twitch account said that the streamer was stupid for allowing the Discord images to show up, and called her a racial epithet (she is a person of color). 

While this was happening, the streamer got flustered and started crying, and one of the mods convinced her to end the stream and delete the VOD immediately. In retrospect, the attack was probably carried out by one person or a coordinated group of people. But at the time it seemed like just a bunch of weird random occurrences, and \#2 and \#3 didn’t seem like obvious attacks.
\end{quote}

Based on this viewer’s description, the group attack was well organized and the role of every attacker was clear. They took advantage of Twitch’s policies and its specific font to shut the stream, where the streamer had no chance to resist. Even after streamers carefully set all the moderation tools to proactively prevent similar incidents, many of them still fell into the trap designed by attackers and were banned by the platform. Even worse, the appealing process was also heavily delayed and alienated the streamers from their communities who could support them, as claimed by many victim streamers. A streamer shared their experience: 

\begin{quote}

I have all the nets in place. Follower-only chat, verified email, high level of moderation, removal of bots. I've been hit twice now. They know how to get around this stuff, if they want to take you out, they will. Each time I had a few guys come in and tell me they were about to get me ban. They spammed a link, I'm guessing to a porno site since that was what I was banned for. The link showed up as \*\*\* since I have links blocked. I deleted the messages with the blocked link but was still ban an hour later each time. I have sent twitch the screenshots of the chat, including DMs to my other social media accounts with 'that's what you get for banning me hahahaha' and 'P\*ssy ass b\*tches like you deserve to die' from multiple random user accounts. Still waiting on my first appeal, let alone my second one.
\end{quote}

According to this streamer, even after setting up all moderation tools with word filtering and deletion, when confronting a group of well-planned attackers, they were still so powerless and got banned. Furthermore, even when the streamer had tenable evidence handed to Twitch, the appealing process was still too slow to provide any practical assistance to the streamer.

\subsubsection{Confusing Hate Raids For Human-engaged Trolls}
There has been confusion between the term hate raids and group trolls. In this context, a troll is someone who is more mischievous than malicious. A streamer might think a hate raid is a group of trolls and not take it seriously, which could be harmful. A small streamer shared their experience of mistaking hate raids as trolls: \emph{``in the beginning, I gave it time. I thought they may chill and it would be fun... but sadly, it started to be racist and annoying though I wasn't really actually annoyed just didn't feel like it was right to leave trolls enjoy freedom in my channel.''} For new and small streamers like them, it was easy to mix the hate raids with group trolls and think that it is normal to have such hateful comments sent by the people whom they think are trolls, and they should learn to live with it to develop their channels and attract more viewers. 

The confusion between trolls and hate raids can prevent streamers from protecting their stream and themselves timely. A mod shared the story of their streamer friend who was threatened with her life safety by hate raiders whom they mistakenly thought were trolls. Initially, they were told to ignore the mass of follow bots. However, after getting lots of hateful comments in the chat, they started to take the measures suggested by Twitch, such as adding bots, banning hateful words, and reporting the accounts. However, it did not seem helpful, and a few months later, the attackers not only posted her personal information, her family’s, and mods’ through chat but also sent a package to her address, which meant they knew where she lived and messaged that they would rape her. \textit{“It’s extremely scary and things have been escalating for months, while we keep trying to ignore them and take the measures suggested by Twitch to stop them.''} In this case, initially they thought that the attackers were just trolls, so they kept ignoring them, which escalated it to an unmanageable level: the personal information of her and the people around her was seriously compromised, and their safety received a huge threat.

\subsection{RQ2: Impacts of Hate Raids on Live Streaming Communities}
\subsubsection{Streamers' Psychological Harm and Diminishing Community}
Many streamers who had experienced hate raids reported that they were scared to stream or were not happy about it. Because they expected either follow bots or hate raids to happen while they were streaming, they were no longer enjoying streaming as before. As a streamer commented: \emph{``Ever since these bots came in to full effect, every time I get a new follower (since I don't get many to begin with) I am IMMEDIATELY suspicious and start searching to see if they're a bot name. I hate doing it but damn it's just concerning now that any time I get a follow, I go into panic mode thinking I'm about to be ass blasted by hate spam.''} Hate raids turned a new following from the most anticipated thing for a small streamer into something to be feared or even trauma. It also made streaming not about enjoying the games and interacting with viewers, but rather concerns about receiving attacks. As a result, streaming became emotionally burdensome and harmed mental health. 

However, if attacked streamers stop streaming, their communities' engagement would decrease, harming their channels, which could severely hurt streamers who stream for a living or an essential source of happiness. A full-time streamer commented, \emph{``My viewer count is dive bombing, since I am unable to make content in my primary game. Everything I have spent years building is crumbling, and there isn’t a thing I can do to stop it, except feed these people, who have chosen to make it their full-time job to make my life a living hell.''} Another user whose streamer friend took a few days off Twitch to avoid hate raids commented, \emph{``The thing is streaming is her livelihood, she doesn't make much but it's enough to live off [of] currently. Not only that but it’s the thing that makes her happy.''} For transgender streamers, many mentioned they were not willing to turn their audio on during streaming because their voices can make them easier to get attacked, despite the fact that not having voice chat would negatively affect their viewership. 


In addition to reducing or even withdrawing streaming, many streamers commented that they were concerned about their safety because some attackers had their personal information and could hurt them in real life. A streamer explained, \emph{``He [the hate raider] said he was a killer and knew information he could only find from my social media account. Like can I go on my Twitch and see the IPs of people who looked on my account? I apologize for the hysteria but I am very much scared right now.''} Such concern for personal safety could seriously influence streamers’ daily lives. A user whose streamer friend had been attacked shared that their friend was not going out alone now and ordered food \emph{``through a delivery gate so that no one can get to her in this manner,''} and was even considering moving: \emph{``She lives alone, everyone is encouraging her to move, and she wants to. But apparently we can’t break the rental contract over a stalker.''} 

Most viewers who witnessed hate raids would leave the stream because it affected viewing experiences. Additionally, some comments pointed out that viewers were concerned that they would be attacked if they stayed in the chatroom: \emph{“viewers leave their stream because they don't want to get doxed and don't want gross spammers sending them sexual whispers (some of the people getting aggressive rapey whispers are minors too which is even worse).”} Some viewers who shared the same identities with the streamer often left the room, fearing being attacked.

\subsubsection{Users' Complaints and Sympathy Toward Twitch's Responses}
Many complained that Twitch did too little to deal with hate raids. Some streamers commented that Twitch had few timely responses. A small streamer shared their experience of reporting attackers they encountered on Twitch \emph{“I reported it to Twitch with a list of the 100 or so names they used, but Twitch didn’t do anything about it, they answered the ticket, but it was like a BS copy-paste kind of response (shrug). It's up to us to moderate ourselves really.”} 

In addition to the insufficient handling of hate raids reports, many users expressed their anger toward Twitch for its slowness in coming up with a systematic solution to the issue. A streamer commented, \emph{``You gotta be kidding, people have been harassed for being part of a minority in twitch for a month, and twitch is only 'working on it', they should already have a solution, it's been a month already, people outside from twitch have come up with temporary solutions but no news from the company itself.''} The streamer pointed out that while users promptly responded by developing temporary solutions to protect themselves and each other from hate raids, Twitch, which should have regulated attacks on its platform, was still so slow and could not provide any solutions or news. 


Streamers who were unsatisfied with Twitch's response conducted the DayOffTwitch campaign and stopped streaming for one single day. While the campaign was designed to warn Twitch, it also influenced many Twitch streamers who were not originally attacked in the hate raids. Some small streamers took advantage of DayOffTwitch by streaming and got many more viewers than in the past because of less competition on that day. However, some other streamers lost followers or even were attacked for streaming. They complained about how the movement deviated from its original meaning of fighting against hate raids and became \textit{``harassing people to protest harassment.''} Likewise, some streamers chose not to stream on the day just to avoid backlash.

However, while many users complained that Twitch was not effectively combating hate raids, some urged people not to expect too much from Twitch. On the one hand, there was a call for more patience with Twitch because it was difficult to devise a solution. \emph{“They are working on it, this isn't as simple as ‘incoming messages that look alike = block’ They need to find a solution that doesn't block normal users, but to only block the bots, and that's not easy.''} Also, since it takes time for Twitch to resolve issues, users should take some personal steps to protect themselves instead of solely relying on Twitch. A user wrote that, \emph{“I’m not insisting Twitch can't do more, but I'm reluctant to place the blame solely on Twitch when we have personal moves we can make. It won't be perfect, but as I've said before, solutions are slow and hard to come by.”} On the other hand, some users explained that it is hard for Twitch and every big game streaming platform to solve the hate raid issue, because \textit{``every time you improve security, you get an influx of new `talent' trying to break the system.''}

\subsubsection{Weighing Other Platforms in Live Streaming Industry}

\paragraph{Users Intended to Leave Twitch but Facing Challenges}

Some users suggested that leaving Twitch and joining other competitors could force Twitch to change: \emph{``Change doesn't happen with overnight threats, it takes sustained throttling of their money avenues. If Twitch suffers, it suffers. Right now, they don't show for a minute that they actually care for their userbase. So if you want them to feel a hit, it needs to be a lasting impression.''} Users believed that only when Twitch suffers long-term financial damage will it be forced to change the situation. However, switching to another platform was difficult for Twitch streamers, and the biggest issue was that their viewers might not be willing to move with them, so they could lose viewership. As a user wrote, \emph{``95\% of streamers aren’t big enough and don’t have the influence enough to bring their audience over to another platform. Even ninja lost almost half his viewers. Twitch knows it. We know it. That's why twitch is the way they are. They know most of us can't afford to start over.''}

\paragraph{Users' Comparison Between Twitch and Its Competitors}

In the discussion among users planning to leave Twitch and join its competitors, YouTube was frequently mentioned and compared to Twitch. Many users pointed out the advantages of YouTube, such as \emph{“better copyright dispute system, larger platform and audience reach, better video and audio quality, unlimited VoD archival,”} and \emph{“better capacity to add the UI elements;”} in terms of solving hate raids. Some users believed that \emph{“YouTube can solve most of those problems easier than what twitch would have to do to get to bitrates and resolutions that compete with YouTube’s.''} However, there were comments on why YouTube might not be a good choice for Twitch users. A user commented that streamers might experience similar harassment and attacks on other live streaming platforms as they have suffered on Twitch. YouTube is \emph{“more strict in terms of monetization''} and has a \emph{“more toxic and often a lot younger”}  viewer community than Twitch. In addition to YouTube, some smaller platforms such as Trovo were also raised as options by some users: \emph{“There is a competitor currently growing, it's called Trovo. It's pretty small but it has been growing at about the same pace Twitch did when it started.”}


\subsection{RQ3: Approaches and Challenges to Combat Hate Raids}
Users have discussed social and technical approaches to combat hate raids, but social approaches are limited and are only discussed by a small group. Most discussions focused on moderation tools and whether they are effective. Not every suggested or existing tool can help solve these issues, but streamers were trying to compile a list of tools and settings to mitigate the impact of hate raids and follow bots. To this end, they were still looking for more effective moderation tools. The tools with descriptions are summarized in \autoref{tools}.

\subsubsection{Limited To No Discussion About Social Approaches}
The social approach discussed mainly focused on how to support streamers with either ``love raids'' or moderation expertise. Some users suggested that streamers and viewers should rally to support those who were victims of hate raids. A user stated, \textit{``I think it would be really cool if we could start a thread of streamers who could use a LOVE raid!”} Another user suggested that \textit{``anyone who is struggling with trolls and haters in a dead chat, they should post in my discord to see who's awake and able to come help guide chat, deal with trolls (in a nice way), and restore your confidence.''} This would help streamers without established audiences since they are less likely to have fans or moderators to help them cope with an attack. This showcases the feeling of community that was widespread throughout these comments. While most people did not discuss this specifically and were not explicit about their support/care for the streamers affected, they showed how the users were concerned about hate raids and follow bots by giving suggestions.


\subsubsection{Proactive Tools That Try to Prevent Attacks}
Twitch chat settings, verification, IP bans, extension review, third-party tools, and improved control over raiding could be helpful before any attacks, by preventing malicious actors from accessing the chat. Verification, if enabled, would require all accounts to be verified before they were able to send messages. Ideally, this would stop attacks involving bots because each bot would have to be verified. A viewer pointed out that the main issue is the cost/benefit ratio for attackers: 

\begin{quote}
    The question for an attacker is essentially: ``is the time/effort for gathering accounts worth the attention I can receive from streamers?'' This cycle currently fulfills itself because streamers at the moment cannot deal with the problem *until* it has happened, so even the act of starting a hate raid, even with the most effective of actions against such will significantly detriment a streamer and bring the attacker satisfaction.
\end{quote}


Essentially, it is important to remember that attackers do this for their enjoyment. If Twitch makes it harder to create and verify bots, then the benefits to the attacker will be less than the cost of the attack. This falls on Twitch to take care of, as streamers cannot prevent the creation of new bots. 

The tools to which streamers have access also have problems. A streamer said, \emph{“I've never seen this feature [requiring verification] work as intended because it's so easy to get around. In fact, you can use one single verified email to create thousands of bot accounts. How is that functional?”} More importantly, as a streamer said, \emph{“almost all streamers refuse to turn on”} the email verification option \emph{“because they don't want to lose potential chatters.”} Streamers did not want to have to choose between engagement and safety from attack. 

Setting one’s chat to only accept messages from those who meet certain standards could also significantly hinder attacks, especially with more stringent rules. Unfortunately, the same problem applies here as well. A viewer said, \emph{“If I go to a channel, and it's followers only.... I leave.”} For smaller streamers, who cannot afford to lose any potential viewers, this makes using strict chat settings untenable. 
If an attacker were IP banned, they would not be able to use alternative accounts to access Twitch, which would prevent what a streamer described as the game of \emph{“Whack a HOSS [bot].”} This was something that many users wanted to see done more often, but there were two main problems. First, IPs are dynamic and can frequently change, so a completely random person could end up being banned while the attacker was still able to access the site. Additionally, a user pointed out that, \emph{“the bots are most certainly using VPNs and randomizing their IP addresses so blocking IPs would be entirely ineffective.”} 

Reviewing extensions for security issues could help prevent people from falling victim to IP grabbers. Extensions add features like subtitles to streams but they often connect to external servers, which can pose a security risk. There was disagreement about whether or not this was a real problem worth spending time on. Those who believed it felt that it was another example of Twitch neglecting its responsibilities. One user said,\emph{``why can't Twitch verify extensions and only let creators use verified extensions? I mean they manually verify emotes right?''} However, there was much confusion about this point. For example, Twitch already does review extensions. One user pointed this out in return: \emph{``And in fact that's what Twitch does. Alice\&Slith had a really hard time getting their extension approved, which delayed their ARG for a few days/weeks.''} This is also corroborated by Twitch's website \footnote{\url{https://dev.twitch.tv/docs/extensions/life-cycle/}}, where the process for developing an extension clearly states that it must be reviewed before it is published. 




Additionally, they pointed out that many worried about IP grabbers didn't really understand what an IP was for, and that there was not much cause for concern since every website someone interacts with can see their IP. Another user explained it by saying \emph{``An IP address is sort of like a license plate number, its rather harmless information for people to have unless they intend to try to leverage everything they know about you against you.''} 

Of course, being able to reject individual raids would help streamers avoid that specific avenue of attack from any suspicious channels, even though most raids don’t actually use the raids feature. Users felt that the options available, which were to either accept raids from everyone, only from friends, or turn off raids entirely, were too restrictive. A streamer said it was like \emph{“a sledgehammer being used for a screwdrivers job”} as raiding was a fundamental way to network and expand one’s audience. Like with verification and chat settings, streamers did not like sacrificing channel growth for safety. For marginalized streamers, their channels allow them to create a community where they can feel safe. Because of this, the prospect of losing engagement can be even worse.  

\subsubsection{Proactive Tools that Prevent the Audience From Being Exposed}
Automod’s word filter system is helpful if a channel has been attacked, as it prevents viewers from seeing messages that include specific words or phrases. Automod can be set to various levels of filtration, as well as banning or allowing specific words and phrases, so one streamer could ban all swearing while another could allow everything except for a word they did not like. Many users recommended it, but one major issue is that \emph{“automod has a ton of issues with lgbtqia+ terms.”} Terms related to the LGBTQ community can be flagged as sexual content, which would mean that a streamer in the community could not take advantage of the filter’s potential as it would ban discussion relevant to LGBTQ issues. In addition, attackers can easily bypass these filters. A black streamer commented that they \emph{“used to get spammed with all kinds of symbols and hate”} including \emph{“getting called a ‘rrigger’.”} As such, many users felt that Automod was an essential but inadequate tool to deal with these issues. 
Some third-party tools include moderation bots with lists of known hate raiders, which can be preemptively blocked. While there are some issues with this - chiefly, that this will always be out of date, as it takes time to add new accounts - this was something that many users appreciated, especially as these are \emph{“simple to set up and just forget it.”} 

\subsubsection{Reactive Tools During an Attack}
Chat settings and third-party tools can help mitigate the effects of an attack while it is occurring. If a hate raid happens, a streamer can either manually or with a third-party tool apply some settings that will stop the attackers from sending messages in the chat. Third-party moderation bots also allow streamers to set up a panic button, a command that executes multiple actions with one button press. A streamer stated, \emph{``I have a Panic button (and an undo panic button) set up, so if for some reason I get some shit going down in chat, I press it, and everything is locked down... Mine will enable emote-only chat, sub-only chat, follow requirement of 5 or 10 min, clear chat, enable slow chat, disable alerts, etc., all with a single button press.''} The benefit is clear, as it is a quick way to deal with an attack. As that streamer said, \emph{“if you’re prepared, you don’t have to worry so much,”} which shows the peace of mind that this tool can provide. 

An important difference between this and simply changing the chat settings from the beginning of the stream is that this does not harm engagement to the same degree. Rather than limiting participation from the beginning of the stream, this allows for a more flexible approach that only affects engagement during the attack. Afterward, the chat can be reverted to normal. This increased flexibility is something that users also wanted in regards to the raids feature itself, which shows that in several areas, Twitch hasn't given users as much control as they regard as necessary. 
A common theme was that while it was good that these third-party tools worked, it was a failure on Twitch’s end that streamers had to rely on third parties. One user said, \emph{“We should have the tools to protect ourselves and a panic button shouldn't be the only tool.”} They were also frustrated that the panic button, which was universally considered an excellent tool, had been created by third parties when they believed this was something \emph{“anyone could of done before.”}




\subsubsection{Reactive Tools After an Attack}

After an attack, bans would help prevent a recurrence from the same person. However, as explained earlier, regular bans and IP bans have drawbacks, so this will not solve the problem entirely. There are also third-party tools, such as serybot, that can identify and automatically ban known bot accounts. Additionally, some third-party tools give streamers the ability to review all followers within a specific time frame. If there is an attack, this allows them to ban all the bots at once. 

On a more communal level, the victim of an attack could report the accounts involved to the developers of a third-party moderation tool, so that they could be added to the list of known attackers. Furthermore, many users wanted to implement multi-channel bans, which would allow communities of streamers to help each other. A person who is a moderator on multiple channels could ban an account, which would prevent that account from being able to access any of the channels that the person moderates. As a streamer described, \emph{“This could be used to prevent users from harassing a group as a whole.”} 
While this step does not help the individual streamer very much, since the attack is over by the time they can do this, it helps the streamer and the community overall better prepared for the next attack. However, this is not very useful against attacks using bots, because it is easy for an attacker to get around bans, and they sometimes use names specifically taunting streamers who have banned them. As one user put it, \emph{``You change your code, and attackers try different tactics. It's an arms race, really.''}

\section{Discussion}



In this study, we extend previous work on online harassment and content moderation and focus on the coordinated group attack in real time in live streaming communities.  Hate raids as a human-bot coordinated  group attack leverages the features of live streaming system to offend marginalize streamers with(out) violating rules. It initially targets marginalized streamers and can extend its targets to any streamers or user groups in live streaming communities, such as streamers who don't join the social media campaign and take advantage of the campaign. The attack pattern can also be generalized to any other platform, especially new platforms with many interactive elements but lack of moderation design. Marginalized streamers suffer from multiple harms, such as psychological harm, community loss, and safety threat. Marginalized streamers applies more technical than social approaches but can not sacrifice the engagement with high-level moderation like big streamers. The lack of effective tools and support to handle hate raids and the sufferings from it pile up marginalized streamers' complaint about and even contest against the platform. Such activities reflect the problems and challenges of the current moderation system design and urgently require new design approaches in the case of crisis management.

\subsection{Affordances of Live Streaming Systems Facilitate Hate Raids}

\subsubsection{Features Abuse Without Violating Moderation Rules}
Attackers utilize the Twitch features, initially designed to build a healthy live streaming environment and promote streamers, in negative ways to harm marginalized streamers. Streamers explicitly mention several features to exacerbate hate raids, making us reflect on the live streaming system design to mitigate feature abuse.  

The identity tag mechanism increases the searchability of marginalized streamers, connects them to people who share the same identities, and promotes equality and their community \cite{Lopez2022ToStreaming}. However, it also provides potential attackers with opportunities to conduct hate raids and increases the scalability of attacks. The live streaming interface leads to asymmetric exposure between attackers and streamers \cite{Uttarapong2021HarassmentNegativity, Zhou2021DesigningCyberbullying} and provides an environment for hate raids. On the one hand, the ``live'' affordances increase streamers' visibility to the public; on the other hand, they expose streamers' appearances and identities to potential attackers. Similarly, the text-based chatroom encourages viewers to engage while hiding attackers' identifiable information with pseudonymous usernames.

Prior work primarily emphasizes how attackers bypass the moderation system to keep breaking rules \cite{Chancellor2016thyghgapp:Communities}. Similarly, the attackers keep generating new usernames to circumvent the block list developed by moderation teams and to send certain hateful words with variants in the chat. Differently, attackers leverage several features initially designed to facilitate live chat interaction. One prominent feature is the ``follow'' notification. Once a user clicks on it and starts following a streamer, the streamer will receive a notification from their side. Attackers trick the notification mechanism to keep ``following/unfollowing'' without breaking any rules to disrupt the interaction in the chat.   

Attackers can also take advantage of the possible offline interaction on Twitch to circumvent active moderation and mode setting (e.g., follow-only mode). It is possible to access a streamer's chat even if the stream is not live, and devoted fans often maintain a small community there. Because offline chats are rarely active, most streamers do not moderate them. There were some anecdotes about attackers going into offline chats, sending hateful messages, and then reporting the streamers for having no moderation. Attackers even utilize Twitch’s terms of service by entrapping streamers to break rules and leverage the replicability of the Internet to secretly record live streaming screens, then report streamers to get them banned.


\subsubsection{Algorithmic Confrontation Between Moderation Tools and Follow Bots}

Moderation tools can actively ban bot accounts, but (1) the simple registration mechanism allows an email to generate multiple accounts, and even bots are allowed to register, and (2) each ban generates a notification in the chatroom. The ease of creating accounts provides fertile ground for attacks on marginalized groups \cite{Massanari2017GamergateTechnocultures}. The easy and quick generation of bots by attackers can use a simple algorithm to incessantly create new bots and send them into the chatroom. Though the tools can actively detect bots with similar usernames (e.g., hoss\_xxx), the mass ban keeps generating notifications in the chat as bots keep joining in. Attackers leverage the simplistic account registration mechanism to confront the moderation algorithm. The synchronicity of the chatroom makes the algorithm confrontation feasible by generating flows of joining and banning notifications in the chatroom. Consequently, conversational messages can easily be lost in flooded notifications. Therefore, hate raids disrupt the conversation in the chatroom and cannot be stopped with available tools; there is no conversational resilience at all in this case \cite{Lambert2022ConversationalEvents}.

\subsubsection{Exploitation of the Platform Governance Structure}

The imbalance between platform-driven and community-driven moderation can create space for potential hate groups to thrive \cite{Seering2019ModeratorAlgorithms}. Twitch's governance structure empowers its community moderation, nonetheless, making it harder to predict and detect hate raids. While proactive moderation tools are more at the platform level to detect video streaming violations, reactive moderation tools are more at the community level to combat chatroom violations. Community moderation like Twitch and Reddit empowers communities to develop their own rules and maintain their communities by themselves \cite{Wohn2019VolunteerExperience,Seering2020ReconsideringModeration}. While it gives users more power in moderation, various rules and norms challenge moderation at scale. No one algorithm/tool can handle multi-level platform governance. The streamer's chat could have enjoyable interaction at the beginning, which straightly passes the platform-level moderation. However, hate raids could suddenly happen in real-time with thousands of bots or with mixed humans and bots. Since hate raids circumvent proactive moderation tools, streamers must take immediate actions manually with their moderators. The large volume of bots and hateful messages overwhelms human labor. Additionally, bot-engaged hate raids might be a problem at the platform level, as they attacked a group of streamers at scale. However, human-engaged hate raids might only be a problem at the community level, depending on community rules and norms. Algorithmic design should consider the situated factors based on the governance structure, leaning a little toward community-level moderation.


\subsection{Marginalized Streamers Endure Multi-level Harms and (Almost) Impossible Trade-offs Between Moderation and Participation}

Prior work has explored marginalized streamers' emotional labor and management and different strategies to handle individual attacks with human moderators and tools \cite{Uttarapong2021HarassmentNegativity}. This study extends this line of research by (1) highlighting other forms of harm caused by hate raids and how live streaming affordances and marginalization amplify these harms (2) and  showing marginalized streamers' struggle to balance moderation and engagement, which is different from moderation strategies to handle individual attacks \cite{Cai2021ModerationCommunities}.

\subsubsection{Multi-level Harms Caused by Hate Raids to Streamers}
Streamers suffer multiple severe harms from hate raids. These harms are not just short-term, but long-lasting \cite{Walker2020MoreSpaces}.
We align the harms with Scheuerman et al.'s harm framework \cite{Scheuerman2021AOnline} to explain different harms that marginalized streamers have suffered from hate raids. 

Marginalized streamers experience mainly three harms: emotional harm, such as the panics of being hate raided and fear of restarting streaming; relational harm, such as streamer-viewer relationship disruption and concern about viewer engagement and community growth; and financial harm, such as viewership loss, community shrink, and subscription decrease. Although few streamers explicitly state physical harm, there is \textit{potential} physical harm, such as safety threats with package delivery to their physical address. 
All these harms are intertwined. For example, safety threat increases psychological burdens with emotional labor and concerns; emotional harm, such as fear of streaming, and relation harm, such as viewership drop, can finally lead to a subscription decrease. 

Several factors amplified the harms: (1) marginalized streamers are the targets, not the bystanders (perspective), (2) attackers highly intend to hurt them (intent), (3) marginalized streamers' harmful experience intensifies harm perception (experience),  (4) hate raids in real-time are human-bot coordinated attacks (scale) and urgent to address with ineffective tools (urgency), (5) marginalized streamers are vulnerable (vulnerability), (6) hate raids can be textual with visual elements (e.g., emoji and memes)  and with the video recording to entrap marginalized streamers (medium), and (7) hate raids happen in the public chatroom (sphere).

\subsubsection{Trade-offs Between Participation and Moderation}
Jiang et al. propose that content moderation is a series of trade-offs regarding moderation actions (e.g., excluding vs. organizing vs. norm-setting), styles (e.g., human vs. automated), philosophies (e.g., nurturing vs. punishing), and values (e.g., community identities) \cite{Jiang2022AModeration}. In this study, we align our findings with relevant trade-offs (actions, styles, and philosophies) to explain how hate raids are challenging to marginalized streamers' communities. Generally, the trade-offs are considered to deal with human-engaged hate raids. However, bot-engaged hate raids sometimes invalidate the trade-off framework and force marginalized streamers to accept the situation.

Regarding moderation actions, the synchronicity and bot-engaged hate raids basically make the trade-off of moderation actions invalid because (1) there is no way to exclude all bots, (2) there is no way to organize content as the instant notifications flowing in the chat, and (3) consequently, there is no way to have meaningful interaction in the chat to set a norm. Regarding moderation styles, the trade-off to deal with bot-engaged hate raids (e.g., mass follow bots with hate messages) is invalid because both humans and automation cannot deal with them effectively. Though automated moderation can constantly capture and ban bots, the algorithmic confrontation between tools and bots disables the interaction in the chat and also makes human labor powerless. The hybrid human and automated moderation might only deal with human-engaged hate raids to some extent. Regarding moderation philosophies, streamers are struggling to make a trade-off between punishing and nurturing. They have expressed concerns between the high-level moderation settings and viewer engagement. The settings (e.g., panic button) can at least mitigate part of hate raids. However, they usually work well with big streamers with a large user base without worrying about losing viewer engagement. Marginalized streamers, usually small streamers and the main target, might be forced to choose participation and community growth over moderation. Additionally, the bot-engaged hate raids might void the trade-off between level of activity and quality of contributions because the restricted moderation might not lower the bot activities at all and consequently increase viewers' contribution. Regarding moderation values, moderators and streamers do not have to make a trade-off because they usually share and maintain community identities.



\subsection{Implications and Recommendations}
Prior work shows that developers notice the flaws in system design to reactively recognize instances of harm to users, backtrack the causes, and fix the mistakes. Recently, Park et al. \cite{Park2022SocialSystems} developed a prototype that can simulate different users and their interactions in either a positive or negative way, to some extent, can automatically identify harmful behaviors caused by the design so that developers can refine the design before deployment. Their research also sheds light on the moderation system design. While affordances indicate the perceived actions associated with the property of the features, the same feature can act in two opposite ways by regular users and potential attackers. 

Combining our findings with prior work, we propose the \textit{moderation-by-design} as a lens when designing new systems and moderation features. \textit{Moderation-by-design} suggests that the mindset of system design should always consider the moderation elements, which is not only the design that facilitates cooperation, but also the mechanism that can potentially prevent abuse of such design. The mechanism should, from a socio-technical perspective, enable stakeholders to adapt and respond quickly through individual or collaborative actions, either proactively or reactively, and minimize the sacrifice of their existing experiences. Developers should consider the negative side when designing the moderation system and better understand the potential abuse of such system, though they have limited direct control over how their designs are enacted \cite{Jones2006ADecade}. With this concept in mind, we provide the following recommendations. Before launching these features, developers should also simulate and test the potential abuse of these features. We propose the tool's design to focus on different stakholders' individual actions and collaboration to combat human-bot coordinated attacks. We clarify that these implications try to mitigate the impact of hate raids on participation; thus, some implications such as simply hiding ``(un)following'' notifications in the chatroom to avoid mass follow bots' impact are not listed because streamers lose the opportunities to interact with new viewers as well.

\subsubsection{Platform Governance With Communication Design} 
\paragraph{Better Channel to Engage with Marginalized Streamers}
Prior work shows that protest users against platforms are more likely to be male and young users \cite{Li2019HowProtest}; this study supplements prior work and shows that marginalized groups can also work as protest users. Twitch is the leading platform in live streaming industry with the possible performance of monopolistic practices \cite{Gillespie2020ExpandingDebates}, invisibly making streamers depend on it for daily needs. Their intention to leave Twitch and join competitors but unable to is in line with previous work that shows that the challenge is the concern of losing community connection \cite{Li2019HowProtest}. This is a reminder to the platform to shift its focus on the big streamers who bring profits to the platform and to actively engage with marginalized streamers. For example, the platform can specifically add a channel to serve marginalized streamers and speed up the reporting and appeal processes.

\paragraph {Better Communication Between the Platform and Users}
The mixed attitudes of the users towards Twitch indicate that Twitch needs better communication with its users to understand the problem and let its users know the necessary information. Although Twitch publishes the transparency report about moderation tools and settings \cite{2021H2Report}, some settings are barely known from the user's perspective. For instance, Twitch developers have clearly stated in their UserVoice \footnote{\url{https://twitch.uservoice.com/}} communities that they have implemented IP Bans already, but the comments showed that this is not well-known by the public. The platform should consider better communicating its roles and actions to the public, at least to marginalized streamers, who always feel excluded and isolated.  

\subsubsection{Implications for System Designer and Developers}

\paragraph{Inclusive and Equitable Moderation Algorithm Design}

Recent work also shows that different user groups consider toxicity differently; for instance, LGBTQ raters are more likely to annotate posts as toxic compared to random raters \cite{Goyal2022IsAnnotation}. Similarly, mitigating online harassment needs to take marginalized users' needs into the platform and moderation system design \cite{Blackwell2017ClassificationHeartMob, Schulenberg2023TowardsReality}. This is in line with Schoenebeck and Blackwell's notion about equality to equity for moderation system design \cite{Schoenebeck2021ReimaginingRepair}. This goal requires developers' input about design goals and rules and community members' values and needs, which require a strong developer and moderator/community member collaboration. However, prior research seemed to focus much on moderator-user interaction, moderator-bot interaction \cite{Cai2021ModerationCommunities,Jhaver2019Human-machineAutomoderator} with little understanding of moderator-developer or community member-developer collaboration. For example, algorithm developers can work closely with LGBTQ+ streamers to update the terms in AutoMod to improve its efficiency and usability, in accordance with user-centered design methodologies throughout the design process \cite{Jones2006ADecade, Hedestig2003FacilitatorsEnvironment} and moderation system development \cite{Chandrasekharan2019Crossmod:Moderators}. A particular space to collect feedback from marginalized streamers could be considered.




\paragraph{Moderation System with Digital Forensics}

Hate raid is not only simple online harassment, but also a kind of cybercrime, as Twitch sued attackers who conducted them \cite{Parrish2021TwitchRaiders}. 
There are many digital forensic tools to perform evidence analysis to identify potential crimes \cite{Javed2022ADirections}. Existing moderation systems can also consider integrating some digital forensics technology to collect, preserve, extract, and report activities conducted by a user, for example, using digital forensic tools to investigate the streamer entrapping cases and tracing the whole process of attackers' behaviors instead of only relying on the image reported by attackers. This way can mitigate the reporting system's abuse and identify the attackers with a chain of evidence.

\paragraph{Engaging Third-party Developers as Ecosystem}
Third-party developers are on the front line and usually encounter and react faster than the platform, as shown in this study. Third parties have already developed some tools to combat hate raids. We argue that third-party developers should be included in the moderation ecosystem \cite{Zuckerman2021WhyEcosystems} and that a mechanism should be implemented to facilitate professional and third-party developer collaboration, possibly by engaging third-party developers in the moderation algorithm design. Furthermore, certain parts of the algorithms should be efficiently utilized and modified by third-party developers. 

\subsubsection{Design to facilitate Streamer-Moderator Collaboration}

\paragraph{Tools to Increase the Visibility and Engagement of Moderators to Streamers}
Streamers and moderators often work as a team to coordinate tasks and manage conflict \cite{Cai2022CoordinationCommunities,Cai2023UnderstandingCommunities}. Sometimes, the streamer lacks active moderators in the chat to provide the necessary help \cite{Cai2022CoordinationCommunities}, especially for new streamers \cite{Zhou2021DesigningCyberbullying}. We recommend tools to support how streamers can identify and need a resource from moderation expertise, for instance, a sidebar with a large available volunteer moderator list on Twitch homepage, showing volunteer moderators' preferences (channel, categories, streamer type, etc.). The list should be large enough to use massive human labor to temporarily join the chat to help the streamer. This perspective argues that streamers can use temporal massive human volunteer labor to combat massive human-bot coordinated attacks.  

\paragraph{Tools to Support Moderation Team Posts in the Chatroom}
Volunteer moderators create large commercial value for the platform, and the platform should show support for their voluntary work with more effective tools \cite{Li2022MeasuringWork}. For example, TrollBuster, as a moderation tool to deal with real-time attacks on Twitter, allows a crisis response team to inundate the victim's Twitter feed with heart-warming and promising tweets to show emotional support to the victim \cite{Ferrier2018TrollBusters:Journalists}. Similarly, a tool should be designed to allow the moderation team to generate massive positive and encouraging messages in the chat when encountering human-engaged hate raids. However, bot-engaged hate raids are different scenarios that the moderation team cannot handle. In these scenarios, something like a conversational bot with positive message post settings should be considered.    

\subsubsection{Design to Facilitate Streamer's Support Seeking and Viewers' Care Giving}

\paragraph {Tools to Better Streamers' Support Seeking From the Same Identity Group}
Prior work suggests that marginalized groups form their communities and safely disclose more about their experience and needs \cite{Haimson2020TransSite, Li2023WeReality}. Marginalized streamers might need more social support from their groups with the same identity on Twitch. Currently, they are sharing their experience granularly (e.g., on Reddit, Discord, Twitter, and other online communities). Twitch is considered a third space for streamer-viewer interaction. Possibly, it can also provide a space for steamers to network and seek different supports \cite{Uttarapong2021HarassmentNegativity} from other streamers, such as instrumental, informational, and emotional support.

\paragraph{Tools to Stimulate Viewers' Positivity to Combat Human-engaged Hate Raids}

Similar to the positivity generator idea by \cite{Ashktorab2016DesigningTeenagers}, some designs can be considered to promote counter-speech to combat human-engaged hate raids. Participating in massive live Twitch chat is less about self-expression and identification, but more about engaging in collective action consistently and continuously \cite{Ford2017ChatChat}. There are tools to promote counter speech from users when the streamer experiences hate speech \cite{Mathew2019ThouSpeech} and to use CAPTCHA to verify human users and simultaneously stimulate positive emotions \cite{Seering2019DesigningBehaviors}. Tools to stimulate viewer engagement are helpful in dealing with human-engaged hate raids, usually just spamming text messages. 

\paragraph{Tools to Crowdsource and Amplify Viewers' Positivity to Mitigate Bot-related Hate Raids}
For bot-engaged or human-bot coordinated hate raids, we recommend a mechanism to facilitate and encourage passive users to use non-text-based communication \cite{Xia2009BallotCommunities} to impact the atmosphere in the chatroom. Therefore, a potential tool should be considered to support crowdsourcing practices of viewers in general, such as crowdsourcing moderation with up and down votes \cite{Lampe2004SlashdotSpace}. Similarly, designers can develop a feature to ensure encouraging messages on the top of the chatroom when the chatroom is full of bots with messages and notifications. This feature may require user-engaged communication tools to participate in content moderation when necessary. For example, a tool can add a stream overlay from the user's perspective to allow all viewers (with passive viewers) to vote \cite{Lessel2017ExpandingStudy} the love raids messages and stick them to the top of the chatroom to amplify the emotional intensity \cite{Luo2020EmotionalEvents} so that the streamer can always see the positive and encouraging words on the top of the chatroom when there is no way to stop the hate raids with constantly flowing messages and notifications.  

\subsection{Limitation and Future Work}
This study has several limitations. First, some quotes are from users in general with no clear roles. We are not sure whether they are streamers or viewers; thus, it might be hard to weigh their significance. Future work should collect data from different affected groups to enrich the depiction of hate raids. Second, the data collection is until February 17th, 2022. Since then, Twitch has been working on some solutions, such as giving steamers control of the `raids' feature \cite{Nightingale2022TwitchHarassment}. There may be more discussion on effective solutions, though our findings suggest that hate raids can be totally irrelevant to the `` raids'' feature.  Future research should try to explore the application of this feature and evaluate its effectiveness to supplement this study. Third, some themes can be explored further, such as how streamers and moderators collaborate to deal with hate raids. Lastly, we only focus on hate raids from the victim's view instead of the attacker's view. Though we menitoned attacks might migrate to other platforms, we know little about the attackers. Future research should explore how hate raiders form their communities and work with bots to attack other communities. This way, we can provide a holistic view about real-time group attacks.

\section{Conclusion}
In this study, we show hate raids as a new form of online harassment that targets marginalized streamers with both human- and bot-engaged attacks and leverages the affordances of live streaming systems to carry out these attacks. These attacks cause multiple severe harms to streamers and force streamers to accept situations with limited trade-offs. Marginalized streamers try more technical approaches rather than social ones, but lack effective tools. We propose moderation-by-design as a philosophy when designing future interactive systems to mitigate potential feature abuse and list suggestions and recommendations to users in live streaming communities.

\begin{acks}
Thank Renkai Ma for the help with data collection. Thank Aashka Patel for the codebook development. This research was funded by  by National Science Foundation (Award No. 1928627).
\end{acks}

\bibliographystyle{ACM-Reference-Format}
\bibliography{references}


\begin{thebibliography}{111}


\ifx \showCODEN    \undefined \def \showCODEN     #1{\unskip}     \fi
\ifx \showDOI      \undefined \def \showDOI       #1{#1}\fi
\ifx \showISBNx    \undefined \def \showISBNx     #1{\unskip}     \fi
\ifx \showISBNxiii \undefined \def \showISBNxiii  #1{\unskip}     \fi
\ifx \showISSN     \undefined \def \showISSN      #1{\unskip}     \fi
\ifx \showLCCN     \undefined \def \showLCCN      #1{\unskip}     \fi
\ifx \shownote     \undefined \def \shownote      #1{#1}          \fi
\ifx \showarticletitle \undefined \def \showarticletitle #1{#1}   \fi
\ifx \showURL      \undefined \def \showURL       {\relax}        \fi
\providecommand\bibfield[2]{#2}
\providecommand\bibinfo[2]{#2}
\providecommand\natexlab[1]{#1}
\providecommand\showeprint[2][]{arXiv:#2}

\bibitem[\protect\citeauthoryear{??}{Def}{[n.d.]}]%
        {DefiningTerms}
 \bibinfo{year}{[n.d.]}\natexlab{}.
\newblock \bibinfo{title}{{Defining “Online Abuse”: A Glossary of Terms}}.
\newblock
\newblock
\urldef\tempurl%
\url{https://onlineharassmentfieldmanual.pen.org/defining-online-harassment-a-glossary-of-terms/}
\showURL{%
\tempurl}


\bibitem[\protect\citeauthoryear{??}{202}{2021}]%
        {2021H2Report}
 \bibinfo{year}{2021}\natexlab{}.
\newblock \bibinfo{title}{{H2 2021 Transparency Report}}.
\newblock
\newblock
\urldef\tempurl%
\url{https://safety.twitch.tv/s/article/H2-2021-Transparency-Report?language=en_US}
\showURL{%
\tempurl}


\bibitem[\protect\citeauthoryear{Ashktorab and Vitak}{Ashktorab and
  Vitak}{2016}]%
        {Ashktorab2016DesigningTeenagers}
\bibfield{author}{\bibinfo{person}{Zahra Ashktorab} {and}
  \bibinfo{person}{Jessica Vitak}.} \bibinfo{year}{2016}\natexlab{}.
\newblock \showarticletitle{{Designing Cyberbullying Mitigation and Prevention
  Solutions through Participatory Design With Teenagers}}. In
  \bibinfo{booktitle}{\emph{Proceedings of the 2016 CHI Conference on Human
  Factors in Computing Systems}}. \bibinfo{publisher}{ACM},
  \bibinfo{address}{New York, NY, USA}, \bibinfo{pages}{3895--3905}.
\newblock
\showISBNx{9781450333627}
\urldef\tempurl%
\url{https://doi.org/10.1145/2858036.2858548}
\showDOI{\tempurl}


\bibitem[\protect\citeauthoryear{Bennett}{Bennett}{2021}]%
        {Bennett2021WhatProtest}
\bibfield{author}{\bibinfo{person}{Connor Bennett}.}
  \bibinfo{year}{2021}\natexlab{}.
\newblock \bibinfo{title}{{What is ‘A Day off Twitch’? Why streamers are
  striking to protest}}.
\newblock
\newblock
\urldef\tempurl%
\url{https://www.dexerto.com/entertainment/what-is-a-day-off-twitch-why-streamers-are-striking-to-protest-1643209/}
\showURL{%
\tempurl}


\bibitem[\protect\citeauthoryear{Blackwell, Dimond, Schoenebeck, and
  Lampe}{Blackwell et~al\mbox{.}}{2017}]%
        {Blackwell2017ClassificationHeartMob}
\bibfield{author}{\bibinfo{person}{Lindsay Blackwell}, \bibinfo{person}{Jill
  Dimond}, \bibinfo{person}{Sarita Schoenebeck}, {and} \bibinfo{person}{Cliff
  Lampe}.} \bibinfo{year}{2017}\natexlab{}.
\newblock \showarticletitle{{Classification and its consequences for online
  harassment: Design insights from HeartMob}}.
\newblock \bibinfo{journal}{\emph{Proceedings of the ACM on Human-Computer
  Interaction}} \bibinfo{volume}{1}, \bibinfo{number}{CSCW} (\bibinfo{date}{11}
  \bibinfo{year}{2017}).
\newblock
\showISSN{25730142}
\urldef\tempurl%
\url{https://doi.org/10.1145/3134659}
\showDOI{\tempurl}


\bibitem[\protect\citeauthoryear{Blackwell, Ellison, Elliott-Deflo, and
  Schwartz}{Blackwell et~al\mbox{.}}{2019}]%
        {Blackwell2019HarassmentGovernance}
\bibfield{author}{\bibinfo{person}{Lindsay Blackwell}, \bibinfo{person}{Nicole
  Ellison}, \bibinfo{person}{Natasha Elliott-Deflo}, {and} \bibinfo{person}{Raz
  Schwartz}.} \bibinfo{year}{2019}\natexlab{}.
\newblock \showarticletitle{{Harassment in Social Virtual Reality: Challenges
  for Platform Governance}}.
\newblock \bibinfo{journal}{\emph{Proceedings of the ACM on Human-Computer
  Interaction}} \bibinfo{volume}{3}, \bibinfo{number}{CSCW} (\bibinfo{date}{11}
  \bibinfo{year}{2019}), \bibinfo{pages}{1--25}.
\newblock
\showISSN{2573-0142}
\urldef\tempurl%
\url{https://doi.org/10.1145/3359202}
\showDOI{\tempurl}


\bibitem[\protect\citeauthoryear{Boyd}{Boyd}{2008}]%
        {Boyd2008TakenPublics}
\bibfield{author}{\bibinfo{person}{Danah~Michele Boyd}.}
  \bibinfo{year}{2008}\natexlab{}.
\newblock \emph{\bibinfo{title}{{Taken out of context: American teen sociality
  in networked publics}}}.
\newblock \bibinfo{thesistype}{Ph.D. Dissertation}. \bibinfo{school}{University
  of California, Berkeley}.
\newblock
\urldef\tempurl%
\url{https://www.proquest.com/openview/9cc930ef134daf46c17434d2992e8251/1?pq-origsite=gscholar&cbl=18750}
\showURL{%
\tempurl}


\bibitem[\protect\citeauthoryear{Braithwaite}{Braithwaite}{2014}]%
        {Braithwaite2014SeriouslyWomen}
\bibfield{author}{\bibinfo{person}{Andrea Braithwaite}.}
  \bibinfo{year}{2014}\natexlab{}.
\newblock \showarticletitle{{‘Seriously, get out’: Feminists on the forums
  and the War(craft) on women}}.
\newblock \bibinfo{journal}{\emph{New Media {\&} Society}}
  \bibinfo{volume}{16}, \bibinfo{number}{5} (\bibinfo{date}{8}
  \bibinfo{year}{2014}), \bibinfo{pages}{703--718}.
\newblock
\showISSN{1461-4448}
\urldef\tempurl%
\url{https://doi.org/10.1177/1461444813489503}
\showDOI{\tempurl}


\bibitem[\protect\citeauthoryear{Brown}{Brown}{2020}]%
        {Brown2020LGBHarassment}
\bibfield{author}{\bibinfo{person}{Anna Brown}.}
  \bibinfo{year}{2020}\natexlab{}.
\newblock \bibinfo{title}{{LGB online daters have positive experiences overall
  but face harassment}}.
\newblock
\newblock
\urldef\tempurl%
\url{https://www.pewresearch.org/fact-tank/2020/04/09/lesbian-gay-and-bisexual-online-daters-report-positive-experiences-but-also-harassment/}
\showURL{%
\tempurl}


\bibitem[\protect\citeauthoryear{Cai and Wohn}{Cai and Wohn}{2019}]%
        {Cai2019CategorizingTwitch}
\bibfield{author}{\bibinfo{person}{Jie Cai} {and} \bibinfo{person}{Donghee~Y.
  Wohn}.} \bibinfo{year}{2019}\natexlab{}.
\newblock \showarticletitle{{Categorizing Live Streaming Moderation Tools: An
  Analysis of Twitch}}.
\newblock \bibinfo{journal}{\emph{International Journal of Interactive
  Communication Systems and Technologies}} \bibinfo{volume}{9},
  \bibinfo{number}{2} (\bibinfo{date}{7} \bibinfo{year}{2019}),
  \bibinfo{pages}{36--50}.
\newblock
\showISSN{2155-4218}
\urldef\tempurl%
\url{https://doi.org/10.4018/IJICST.2019070103}
\showDOI{\tempurl}


\bibitem[\protect\citeauthoryear{Cai and Wohn}{Cai and Wohn}{2021}]%
        {Cai2021AfterCommunities}
\bibfield{author}{\bibinfo{person}{Jie Cai} {and}
  \bibinfo{person}{Donghee~Yvette Wohn}.} \bibinfo{year}{2021}\natexlab{}.
\newblock \showarticletitle{{After Violation But Before Sanction: Understanding
  Volunteer Moderators' Profiling Processes Toward Violators in Live Streaming
  Communities}}.
\newblock \bibinfo{journal}{\emph{Proceedings of the ACM on Human-Computer
  Interaction}} \bibinfo{volume}{5}, \bibinfo{number}{CSCW2}
  (\bibinfo{date}{10} \bibinfo{year}{2021}), \bibinfo{pages}{1--25}.
\newblock
\showISSN{2573-0142}
\urldef\tempurl%
\url{https://doi.org/10.1145/3479554}
\showDOI{\tempurl}


\bibitem[\protect\citeauthoryear{Cai and Wohn}{Cai and Wohn}{2022}]%
        {Cai2022CoordinationCommunities}
\bibfield{author}{\bibinfo{person}{Jie Cai} {and}
  \bibinfo{person}{Donghee~Yvette Wohn}.} \bibinfo{year}{2022}\natexlab{}.
\newblock \showarticletitle{{Coordination and Collaboration: How do Volunteer
  Moderators Work as a Team in Live Streaming Communities?}}. In
  \bibinfo{booktitle}{\emph{CHI Conference on Human Factors in Computing
  Systems}}. \bibinfo{publisher}{ACM}, \bibinfo{address}{New York, NY, USA},
  \bibinfo{pages}{1--14}.
\newblock
\showISBNx{9781450391573}
\urldef\tempurl%
\url{https://doi.org/10.1145/3491102.3517628}
\showDOI{\tempurl}


\bibitem[\protect\citeauthoryear{Cai, Wohn, and Almoqbel}{Cai
  et~al\mbox{.}}{2021}]%
        {Cai2021ModerationCommunities}
\bibfield{author}{\bibinfo{person}{Jie Cai}, \bibinfo{person}{Donghee~Y. Wohn},
  {and} \bibinfo{person}{Mashael Almoqbel}.} \bibinfo{year}{2021}\natexlab{}.
\newblock \showarticletitle{{Moderation Visibility: Mapping the Strategies of
  Volunteer Moderators in Live Streaming Micro Communities}}. In
  \bibinfo{booktitle}{\emph{Proceedings of ACM International Conference on
  Interactive Media Experiences}}. \bibinfo{pages}{61--72}.
\newblock
\showISBNx{9781450383899}
\urldef\tempurl%
\url{https://doi.org/10.1145/3452918.3458796}
\showDOI{\tempurl}


\bibitem[\protect\citeauthoryear{Cai and Yvette~Wohn}{Cai and
  Yvette~Wohn}{2023}]%
        {Cai2023UnderstandingCommunities}
\bibfield{author}{\bibinfo{person}{Jie Cai} {and} \bibinfo{person}{Donghee
  Yvette~Wohn}.} \bibinfo{year}{2023}\natexlab{}.
\newblock \showarticletitle{{Understanding Moderators' Conflict and Conflict
  Management Strategies with Streamers in Live Streaming Communities;
  Understanding Moderators' Conflict and Conflict Management Strategies with
  Streamers in Live Streaming Communities}}. In
  \bibinfo{booktitle}{\emph{Proceedings of the 2023 CHI Conference on Human
  Factors in Computing Systems (CHI ’23)}}. \bibinfo{publisher}{ACM},
  \bibinfo{address}{Hamburg, Germany}, \bibinfo{pages}{1--12}.
\newblock
\showISBNx{9781450394215}
\urldef\tempurl%
\url{https://doi.org/10.1145/3544548.3580982}
\showDOI{\tempurl}


\bibitem[\protect\citeauthoryear{Chancellor, Hu, and De~Choudhury}{Chancellor
  et~al\mbox{.}}{2018}]%
        {Chancellor2018NormsCommunities}
\bibfield{author}{\bibinfo{person}{Stevie Chancellor}, \bibinfo{person}{Andrea
  Hu}, {and} \bibinfo{person}{Munmun De~Choudhury}.}
  \bibinfo{year}{2018}\natexlab{}.
\newblock \showarticletitle{{Norms matter: Contrasting social support around
  behavior change in online weight loss communities}}.
\newblock \bibinfo{journal}{\emph{Conference on Human Factors in Computing
  Systems - Proceedings}}  \bibinfo{volume}{2018-April} (\bibinfo{date}{4}
  \bibinfo{year}{2018}).
\newblock
\showISBNx{9781450356206}
\urldef\tempurl%
\url{https://doi.org/10.1145/3173574.3174240}
\showDOI{\tempurl}


\bibitem[\protect\citeauthoryear{Chancellor, Pater, Clear, Gilbert, and
  De~Choudhury}{Chancellor et~al\mbox{.}}{2016}]%
        {Chancellor2016thyghgapp:Communities}
\bibfield{author}{\bibinfo{person}{Stevie Chancellor},
  \bibinfo{person}{Jessica~Annette Pater}, \bibinfo{person}{Trustin Clear},
  \bibinfo{person}{Eric Gilbert}, {and} \bibinfo{person}{Munmun De~Choudhury}.}
  \bibinfo{year}{2016}\natexlab{}.
\newblock \showarticletitle{{{\#}thyghgapp: Instagram Content Moderation and
  Lexical Variation in Pro-Eating Disorder Communities}}. In
  \bibinfo{booktitle}{\emph{Proceedings of the ACM Conference on
  Computer-Supported Cooperative Work and Social Computing}}.
  \bibinfo{pages}{1201--1213}.
\newblock
\showISBNx{9781450335928}
\urldef\tempurl%
\url{https://doi.org/10.1145/2818048.2819963}
\showDOI{\tempurl}


\bibitem[\protect\citeauthoryear{Chandrasekharan, Gandhi, Mustelier, and
  Gilbert}{Chandrasekharan et~al\mbox{.}}{2019}]%
        {Chandrasekharan2019Crossmod:Moderators}
\bibfield{author}{\bibinfo{person}{Eshwar Chandrasekharan},
  \bibinfo{person}{Chaitrali Gandhi}, \bibinfo{person}{Matthew~Wortley
  Mustelier}, {and} \bibinfo{person}{Eric Gilbert}.}
  \bibinfo{year}{2019}\natexlab{}.
\newblock \showarticletitle{{Crossmod: A Cross-Community Learning-based System
  to Assist Reddit Moderators}}.
\newblock \bibinfo{journal}{\emph{Proceedings of the ACM on Human-Computer
  Interaction}} \bibinfo{volume}{3}, \bibinfo{number}{CSCW}
  (\bibinfo{year}{2019}), \bibinfo{pages}{1--30}.
\newblock
\showISSN{2573-0142}
\urldef\tempurl%
\url{https://doi.org/10.1145/3359276}
\showDOI{\tempurl}


\bibitem[\protect\citeauthoryear{Chandrasekharan, Samory, Srinivasan, and
  Gilbert}{Chandrasekharan et~al\mbox{.}}{2017}]%
        {Chandrasekharan2017TheData}
\bibfield{author}{\bibinfo{person}{Eshwar Chandrasekharan},
  \bibinfo{person}{Mattia Samory}, \bibinfo{person}{Anirudh Srinivasan}, {and}
  \bibinfo{person}{Eric Gilbert}.} \bibinfo{year}{2017}\natexlab{}.
\newblock \showarticletitle{{The bag of communities: Identifying abusive
  behavior online with preexisting internet data}}.
\newblock \bibinfo{journal}{\emph{Conference on Human Factors in Computing
  Systems - Proceedings}}  \bibinfo{volume}{2017-May} (\bibinfo{date}{5}
  \bibinfo{year}{2017}), \bibinfo{pages}{3175--3187}.
\newblock
\showISBNx{9781450346559}
\urldef\tempurl%
\url{https://doi.org/10.1145/3025453.3026018}
\showDOI{\tempurl}


\bibitem[\protect\citeauthoryear{Chatzakou, Kourtellis, Blackburn,
  De~Cristofaro, Stringhini, and Vakali}{Chatzakou et~al\mbox{.}}{2017a}]%
        {Chatzakou2017HateTwitter}
\bibfield{author}{\bibinfo{person}{Despoina Chatzakou},
  \bibinfo{person}{Nicolas Kourtellis}, \bibinfo{person}{Jeremy Blackburn},
  \bibinfo{person}{Emiliano De~Cristofaro}, \bibinfo{person}{Gianluca
  Stringhini}, {and} \bibinfo{person}{Athena Vakali}.}
  \bibinfo{year}{2017}\natexlab{a}.
\newblock \showarticletitle{{Hate is not Binary: Studying Abusive Behavior of
  {\textbackslash}{\#}GamerGate on Twitter}}. In
  \bibinfo{booktitle}{\emph{Proceedings of the 28th ACM Conference on Hypertext
  and Social Media}}. \bibinfo{publisher}{ACM}, \bibinfo{address}{New York, NY,
  USA}, \bibinfo{pages}{65--74}.
\newblock
\showISBNx{9781450347082}
\urldef\tempurl%
\url{https://doi.org/10.1145/3078714.3078721}
\showDOI{\tempurl}


\bibitem[\protect\citeauthoryear{Chatzakou, Kourtellis, Blackburn,
  De~Cristofaro, Stringhini, and Vakali}{Chatzakou et~al\mbox{.}}{2017b}]%
        {Chatzakou2017MeasuringBullying}
\bibfield{author}{\bibinfo{person}{Despoina Chatzakou},
  \bibinfo{person}{Nicolas Kourtellis}, \bibinfo{person}{Jeremy Blackburn},
  \bibinfo{person}{Emiliano De~Cristofaro}, \bibinfo{person}{Gianluca
  Stringhini}, {and} \bibinfo{person}{Athena Vakali}.}
  \bibinfo{year}{2017}\natexlab{b}.
\newblock \showarticletitle{{Measuring {\textbackslash}{\#}GamerGate: A Tale of
  Hate, Sexism, and Bullying}}. In \bibinfo{booktitle}{\emph{Proceedings of the
  26th International Conference on World Wide Web Companion}}.
  \bibinfo{publisher}{ACM Press}, \bibinfo{address}{New York, New York, USA},
  \bibinfo{pages}{1285--1290}.
\newblock
\showISBNx{9781450349147}
\urldef\tempurl%
\url{https://doi.org/10.1145/3041021.3053890}
\showDOI{\tempurl}


\bibitem[\protect\citeauthoryear{Consalvo}{Consalvo}{2012}]%
        {Consalvo2012ConfrontingScholars}
\bibfield{author}{\bibinfo{person}{Mia Consalvo}.}
  \bibinfo{year}{2012}\natexlab{}.
\newblock \showarticletitle{{Confronting Toxic Gamer Culture: A Challenge for
  Feminist Game Studies Scholars}}.
\newblock \bibinfo{journal}{\emph{Journal of Gender, New Media, and
  Technology}} \bibinfo{number}{1} (\bibinfo{date}{11} \bibinfo{year}{2012}),
  \bibinfo{pages}{1--11}.
\newblock
\showISSN{2325-0496}
\urldef\tempurl%
\url{https://doi.org/10.7264/N33X84KH}
\showDOI{\tempurl}


\bibitem[\protect\citeauthoryear{Cote}{Cote}{2017}]%
        {cote2017can}
\bibfield{author}{\bibinfo{person}{Amanda~C.e Cote}.}
  \bibinfo{year}{2017}\natexlab{}.
\newblock \showarticletitle{{"I Can Defend Myself": Women's Strategies for
  Coping with Harassment while Gaming Online}}.
\newblock \bibinfo{journal}{\emph{Games and Culture}} \bibinfo{volume}{12},
  \bibinfo{number}{2} (\bibinfo{year}{2017}), \bibinfo{pages}{136--155}.
\newblock
\showISSN{15554139}
\urldef\tempurl%
\url{https://doi.org/10.1177/1555412015587603}
\showDOI{\tempurl}


\bibitem[\protect\citeauthoryear{Das, Dang, and Lease}{Das
  et~al\mbox{.}}{2020}]%
        {Das2020FastContent}
\bibfield{author}{\bibinfo{person}{Anubrata Das}, \bibinfo{person}{Brandon
  Dang}, {and} \bibinfo{person}{Matthew Lease}.}
  \bibinfo{year}{2020}\natexlab{}.
\newblock \showarticletitle{{Fast, Accurate, and Healthier: Interactive
  Blurring Helps Moderators Reduce Exposure to Harmful Content}}.
\newblock \bibinfo{journal}{\emph{Proceedings of the AAAI Conference on Human
  Computation and Crowdsourcing}} \bibinfo{volume}{8}, \bibinfo{number}{1}
  (\bibinfo{year}{2020}), \bibinfo{pages}{33--42}.
\newblock
\urldef\tempurl%
\url{https://ojs.aaai.org/index.php/HCOMP/article/view/7461}
\showURL{%
\tempurl}


\bibitem[\protect\citeauthoryear{Dosono and Semaan}{Dosono and Semaan}{2019}]%
        {Dosono2019ModerationCommunities}
\bibfield{author}{\bibinfo{person}{Bryan Dosono} {and} \bibinfo{person}{Bryan
  Semaan}.} \bibinfo{year}{2019}\natexlab{}.
\newblock \showarticletitle{{Moderation Practices as Emotional Labor in
  Sustaining Online Communities}}. In \bibinfo{booktitle}{\emph{Proceedings of
  the ACM Conference on Human Factors in Computing Systems}}.
  \bibinfo{pages}{1--13}.
\newblock
\showISBNx{9781450359702}
\urldef\tempurl%
\url{https://doi.org/10.1145/3290605.3300372}
\showDOI{\tempurl}


\bibitem[\protect\citeauthoryear{Drosos and Guo}{Drosos and Guo}{2022}]%
        {Drosos2022TheOpportunities}
\bibfield{author}{\bibinfo{person}{Ian Drosos} {and} \bibinfo{person}{Philip~J
  Guo}.} \bibinfo{year}{2022}\natexlab{}.
\newblock \showarticletitle{{The Design Space of Livestreaming Equipment
  Setups: Tradeoffs, Challenges, and Opportunities}}. In
  \bibinfo{booktitle}{\emph{Designing Interactive Systems Conference}}.
  \bibinfo{publisher}{ACM}, \bibinfo{address}{New York, NY, USA},
  \bibinfo{pages}{835--848}.
\newblock
\showISBNx{9781450393584}
\urldef\tempurl%
\url{https://doi.org/10.1145/3532106.3533489}
\showDOI{\tempurl}


\bibitem[\protect\citeauthoryear{Fereday and Muir-Cochrane}{Fereday and
  Muir-Cochrane}{2006}]%
        {Fereday2006DemonstratingDevelopment}
\bibfield{author}{\bibinfo{person}{Jennifer Fereday} {and}
  \bibinfo{person}{Eimear Muir-Cochrane}.} \bibinfo{year}{2006}\natexlab{}.
\newblock \showarticletitle{{Demonstrating Rigor Using Thematic Analysis: A
  Hybrid Approach of Inductive and Deductive Coding and Theme Development}}.
\newblock \bibinfo{journal}{\emph{International Journal of Qualitative
  Methods}} \bibinfo{volume}{5}, \bibinfo{number}{1} (\bibinfo{date}{3}
  \bibinfo{year}{2006}), \bibinfo{pages}{80--92}.
\newblock
\showISSN{1609-4069}
\urldef\tempurl%
\url{https://doi.org/10.1177/160940690600500107}
\showDOI{\tempurl}


\bibitem[\protect\citeauthoryear{Ferrier and Garud-Patkar}{Ferrier and
  Garud-Patkar}{2018}]%
        {Ferrier2018TrollBusters:Journalists}
\bibfield{author}{\bibinfo{person}{Michelle Ferrier} {and}
  \bibinfo{person}{Nisha Garud-Patkar}.} \bibinfo{year}{2018}\natexlab{}.
\newblock \showarticletitle{{TrollBusters: Fighting Online Harassment of Women
  Journalists}}.
\newblock In \bibinfo{booktitle}{\emph{Mediating Misogyny}}.
  \bibinfo{publisher}{Springer International Publishing},
  \bibinfo{address}{Cham}, \bibinfo{pages}{311--332}.
\newblock
\urldef\tempurl%
\url{https://doi.org/10.1007/978-3-319-72917-6{\_}16}
\showDOI{\tempurl}


\bibitem[\protect\citeauthoryear{Feuston, Taylor, and Piper}{Feuston
  et~al\mbox{.}}{2020}]%
        {Feuston2020ConformityModeration}
\bibfield{author}{\bibinfo{person}{Jessica~L. Feuston},
  \bibinfo{person}{Alex~S. Taylor}, {and} \bibinfo{person}{Anne~Marie Piper}.}
  \bibinfo{year}{2020}\natexlab{}.
\newblock \showarticletitle{{Conformity of Eating Disorders through Content
  Moderation}}.
\newblock \bibinfo{journal}{\emph{Proceedings of the ACM on Human-Computer
  Interaction}} \bibinfo{volume}{4}, \bibinfo{number}{CSCW1} (\bibinfo{date}{5}
  \bibinfo{year}{2020}), \bibinfo{pages}{1--28}.
\newblock
\showISSN{25730142}
\urldef\tempurl%
\url{https://doi.org/10.1145/3392845}
\showDOI{\tempurl}


\bibitem[\protect\citeauthoryear{Ford, Gardner, Horgan, Liu, Tsaasan, Nardi,
  and Rickman}{Ford et~al\mbox{.}}{2017}]%
        {Ford2017ChatChat}
\bibfield{author}{\bibinfo{person}{Colin Ford}, \bibinfo{person}{Dan Gardner},
  \bibinfo{person}{Leah~Elaine Horgan}, \bibinfo{person}{Calvin Liu},
  \bibinfo{person}{a.~m. Tsaasan}, \bibinfo{person}{Bonnie Nardi}, {and}
  \bibinfo{person}{Jordan Rickman}.} \bibinfo{year}{2017}\natexlab{}.
\newblock \showarticletitle{{Chat Speed OP PogChamp: Practices of Coherence in
  Massive Twitch Chat}}. In \bibinfo{booktitle}{\emph{Proceedings of the 2017
  CHI Conference Extended Abstracts on Human Factors in Computing Systems}}.
  \bibinfo{publisher}{ACM}, \bibinfo{address}{New York, NY, USA},
  \bibinfo{pages}{858--871}.
\newblock
\showISBNx{9781450346566}
\urldef\tempurl%
\url{https://doi.org/10.1145/3027063.3052765}
\showDOI{\tempurl}


\bibitem[\protect\citeauthoryear{Fox and Tang}{Fox and Tang}{2017}]%
        {Fox2017WomensStrategies}
\bibfield{author}{\bibinfo{person}{Jesse Fox} {and} \bibinfo{person}{Wai~Yen
  Tang}.} \bibinfo{year}{2017}\natexlab{}.
\newblock \showarticletitle{{Women’s experiences with general and sexual
  harassment in online video games: Rumination, organizational responsiveness,
  withdrawal, and coping strategies}}.
\newblock \bibinfo{journal}{\emph{New Media and Society}} \bibinfo{volume}{19},
  \bibinfo{number}{8} (\bibinfo{date}{3} \bibinfo{year}{2017}),
  \bibinfo{pages}{1290--1307}.
\newblock
\showISSN{14617315}
\urldef\tempurl%
\url{https://doi.org/10.1177/1461444816635778}
\showDOI{\tempurl}


\bibitem[\protect\citeauthoryear{Friedman* and Resnick}{Friedman* and
  Resnick}{2001}]%
        {Friedman2001ThePseudonyms}
\bibfield{author}{\bibinfo{person}{Eric~J. Friedman*} {and}
  \bibinfo{person}{Paul Resnick}.} \bibinfo{year}{2001}\natexlab{}.
\newblock \showarticletitle{{The Social Cost of Cheap Pseudonyms}}.
\newblock \bibinfo{journal}{\emph{Journal of Economics <html{\_}ent
  glyph="@amp;" ascii="{\&}amp;"/> Management Strategy}} \bibinfo{volume}{10},
  \bibinfo{number}{2} (\bibinfo{date}{6} \bibinfo{year}{2001}),
  \bibinfo{pages}{173--199}.
\newblock
\showISSN{1058-6407}
\urldef\tempurl%
\url{https://doi.org/10.1111/j.1430-9134.2001.00173.x}
\showDOI{\tempurl}


\bibitem[\protect\citeauthoryear{Gillespie}{Gillespie}{2020}]%
        {Gillespie2020ContentScale}
\bibfield{author}{\bibinfo{person}{Tarleton Gillespie}.}
  \bibinfo{year}{2020}\natexlab{}.
\newblock \showarticletitle{{Content moderation, AI, and the question of
  scale}}.
\newblock \bibinfo{journal}{\emph{Big Data {\&} Society}} \bibinfo{volume}{7},
  \bibinfo{number}{2} (\bibinfo{date}{7} \bibinfo{year}{2020}),
  \bibinfo{pages}{1--5}.
\newblock
\showISSN{2053-9517}
\urldef\tempurl%
\url{https://doi.org/10.1177/2053951720943234}
\showDOI{\tempurl}


\bibitem[\protect\citeauthoryear{Gillespie, Aufderheide, Carmi, Gerrard, Gorwa,
  Matamoros-Fern{\'{a}}ndez, Roberts, Sinnreich, and West}{Gillespie
  et~al\mbox{.}}{2020}]%
        {Gillespie2020ExpandingDebates}
\bibfield{author}{\bibinfo{person}{Tarleton Gillespie},
  \bibinfo{person}{Patricia Aufderheide}, \bibinfo{person}{Elinor Carmi},
  \bibinfo{person}{Ysabel Gerrard}, \bibinfo{person}{Robert Gorwa},
  \bibinfo{person}{Ariadna Matamoros-Fern{\'{a}}ndez},
  \bibinfo{person}{Sarah~T. Roberts}, \bibinfo{person}{Aram Sinnreich}, {and}
  \bibinfo{person}{Sarah~Myers West}.} \bibinfo{year}{2020}\natexlab{}.
\newblock \showarticletitle{{Expanding the debate about content moderation:
  Scholarly research agendas for the coming policy debates}}.
\newblock \bibinfo{journal}{\emph{Internet Policy Review}} \bibinfo{volume}{9},
  \bibinfo{number}{4} (\bibinfo{year}{2020}), \bibinfo{pages}{1--29}.
\newblock
\urldef\tempurl%
\url{https://doi.org/10.14763/2020.4.1512}
\showDOI{\tempurl}


\bibitem[\protect\citeauthoryear{Goyal, Kivlichan, Rosen, and Vasserman}{Goyal
  et~al\mbox{.}}{2022a}]%
        {Goyal2022IsAnnotation}
\bibfield{author}{\bibinfo{person}{Nitesh Goyal}, \bibinfo{person}{Ian
  Kivlichan}, \bibinfo{person}{Rachel Rosen}, {and} \bibinfo{person}{Lucy
  Vasserman}.} \bibinfo{year}{2022}\natexlab{a}.
\newblock \showarticletitle{{Is Your Toxicity My Toxicity? Exploring the Impact
  of Rater Identity on Toxicity Annotation}}.
\newblock \bibinfo{journal}{\emph{Proceedings of the ACM on Human-Computer
  Interaction}} \bibinfo{number}{CSCW} (\bibinfo{date}{5}
  \bibinfo{year}{2022}).
\newblock
\urldef\tempurl%
\url{https://doi.org/10.1145/nnnnnnn.nnnnnnn}
\showDOI{\tempurl}


\bibitem[\protect\citeauthoryear{Goyal, Park, and Vasserman}{Goyal
  et~al\mbox{.}}{2022b}]%
        {Goyal2022YouHarassment}
\bibfield{author}{\bibinfo{person}{Nitesh Goyal}, \bibinfo{person}{Leslie
  Park}, {and} \bibinfo{person}{Lucy Vasserman}.}
  \bibinfo{year}{2022}\natexlab{b}.
\newblock \showarticletitle{{”You have to prove the threat is real”:
  Understanding the needs of Female Journalists and Activists to Document and
  Report Online Harassment}}. In \bibinfo{booktitle}{\emph{CHI Conference on
  Human Factors in Computing Systems}}. \bibinfo{publisher}{ACM},
  \bibinfo{address}{New York, NY, USA}, \bibinfo{pages}{1--17}.
\newblock
\showISBNx{9781450391573}
\urldef\tempurl%
\url{https://doi.org/10.1145/3491102.3517517}
\showDOI{\tempurl}


\bibitem[\protect\citeauthoryear{Grayson}{Grayson}{2022}]%
        {Grayson2022HowFacebook}
\bibfield{author}{\bibinfo{person}{Nathan Grayson}.}
  \bibinfo{year}{2022}\natexlab{}.
\newblock \bibinfo{title}{{How Twitch took down the Buffalo shooter’s stream
  faster than Facebook}}.
\newblock
\newblock
\urldef\tempurl%
\url{https://www.washingtonpost.com/video-games/2022/05/20/twitch-buffalo-shooter-facebook-nypd-interview/}
\showURL{%
\tempurl}


\bibitem[\protect\citeauthoryear{Grimmelmann}{Grimmelmann}{2015}]%
        {Grimmelmann2015}
\bibfield{author}{\bibinfo{person}{James Grimmelmann}.}
  \bibinfo{year}{2015}\natexlab{}.
\newblock \showarticletitle{{The Virtues of Moderation}}.
\newblock \bibinfo{journal}{\emph{Yale Journal of Law and Technology}}
  \bibinfo{volume}{17}, \bibinfo{number}{1} (\bibinfo{year}{2015}),
  \bibinfo{pages}{68}.
\newblock
\urldef\tempurl%
\url{https://digitalcommons.law.yale.edu/yjolt/vol17/iss1/2}
\showURL{%
\tempurl}


\bibitem[\protect\citeauthoryear{Haimson, Buss, Weinger, Starks, Gorrell, and
  Baron}{Haimson et~al\mbox{.}}{2020}]%
        {Haimson2020TransSite}
\bibfield{author}{\bibinfo{person}{Oliver~L. Haimson}, \bibinfo{person}{Justin
  Buss}, \bibinfo{person}{Zu Weinger}, \bibinfo{person}{Denny~L. Starks},
  \bibinfo{person}{Dykee Gorrell}, {and} \bibinfo{person}{Briar~Sweetbriar
  Baron}.} \bibinfo{year}{2020}\natexlab{}.
\newblock \showarticletitle{{Trans Time: Safety, Privacy, and ContentWarnings
  on a Transgender-Specific Social Media Site}}.
\newblock \bibinfo{journal}{\emph{Proceedings of the ACM on Human-Computer
  Interaction}} \bibinfo{volume}{4}, \bibinfo{number}{CSCW2}
  (\bibinfo{date}{10} \bibinfo{year}{2020}).
\newblock
\showISSN{25730142}
\urldef\tempurl%
\url{https://doi.org/10.1145/3415195}
\showDOI{\tempurl}


\bibitem[\protect\citeauthoryear{Haimson, Delmonaco, Nie, and Wegner}{Haimson
  et~al\mbox{.}}{2021}]%
        {Haimson2021DisproportionateAreas}
\bibfield{author}{\bibinfo{person}{Oliver~L. Haimson}, \bibinfo{person}{Daniel
  Delmonaco}, \bibinfo{person}{Peipei Nie}, {and} \bibinfo{person}{Andrea
  Wegner}.} \bibinfo{year}{2021}\natexlab{}.
\newblock \showarticletitle{{Disproportionate Removals and Differing Content
  Moderation Experiences for Conservative, Transgender, and Black Social Media
  Users: Marginalization and Moderation Gray Areas}}.
\newblock \bibinfo{journal}{\emph{Proceedings of the ACM on Human-Computer
  Interaction}} \bibinfo{volume}{5}, \bibinfo{number}{CSCW2}
  (\bibinfo{date}{10} \bibinfo{year}{2021}), \bibinfo{pages}{1--35}.
\newblock
\showISSN{2573-0142}
\urldef\tempurl%
\url{https://doi.org/10.1145/3479610}
\showDOI{\tempurl}


\bibitem[\protect\citeauthoryear{Hamilton, Garretson, and Kerne}{Hamilton
  et~al\mbox{.}}{2014}]%
        {Hamilton2014StreamingMedia}
\bibfield{author}{\bibinfo{person}{William~A. Hamilton},
  \bibinfo{person}{Oliver Garretson}, {and} \bibinfo{person}{Andruid Kerne}.}
  \bibinfo{year}{2014}\natexlab{}.
\newblock \showarticletitle{{Streaming on twitch: Fostering participatory
  communities of play within live mixed media}}. In
  \bibinfo{booktitle}{\emph{Conference on Human Factors in Computing Systems -
  Proceedings}}. \bibinfo{publisher}{ACM}, \bibinfo{address}{New York, NY,
  USA}, \bibinfo{pages}{1315--1324}.
\newblock
\showISBNx{9781450324731}
\urldef\tempurl%
\url{https://doi.org/10.1145/2556288.2557048}
\showDOI{\tempurl}


\bibitem[\protect\citeauthoryear{Hedestig and Kaptelinin}{Hedestig and
  Kaptelinin}{2003}]%
        {Hedestig2003FacilitatorsEnvironment}
\bibfield{author}{\bibinfo{person}{U. Hedestig} {and} \bibinfo{person}{V.
  Kaptelinin}.} \bibinfo{year}{2003}\natexlab{}.
\newblock \showarticletitle{{Facilitator's invisible expertise and
  supra-situational activities in a telelearning environment}}. In
  \bibinfo{booktitle}{\emph{Proceedings of the 36th Annual Hawaii International
  Conference on System Sciences}}. \bibinfo{publisher}{IEEE},
  \bibinfo{pages}{10}.
\newblock
\showISBNx{0769518745}
\urldef\tempurl%
\url{https://doi.org/10.1109/HICSS.2003.1173637}
\showDOI{\tempurl}


\bibitem[\protect\citeauthoryear{Horsman}{Horsman}{2018}]%
        {Horsman2018APeriscope}
\bibfield{author}{\bibinfo{person}{Graeme Horsman}.}
  \bibinfo{year}{2018}\natexlab{}.
\newblock \showarticletitle{{A forensic examination of the technical and legal
  challenges surrounding the investigation of child abuse on live streaming
  platforms: A case study on Periscope}}.
\newblock \bibinfo{journal}{\emph{Journal of Information Security and
  Applications}}  \bibinfo{volume}{42} (\bibinfo{date}{10}
  \bibinfo{year}{2018}), \bibinfo{pages}{107--117}.
\newblock
\showISSN{22142126}
\urldef\tempurl%
\url{https://doi.org/10.1016/j.jisa.2018.07.009}
\showDOI{\tempurl}


\bibitem[\protect\citeauthoryear{Hutchby}{Hutchby}{2001}]%
        {Hutchby2001TechnologiesAffordances}
\bibfield{author}{\bibinfo{person}{Ian Hutchby}.}
  \bibinfo{year}{2001}\natexlab{}.
\newblock \showarticletitle{{Technologies, texts and affordances}}.
\newblock \bibinfo{journal}{\emph{Sociology}} \bibinfo{volume}{35},
  \bibinfo{number}{2} (\bibinfo{date}{5} \bibinfo{year}{2001}),
  \bibinfo{pages}{441--456}.
\newblock
\showISSN{00380385}
\urldef\tempurl%
\url{https://doi.org/10.1017/S0038038501000219}
\showDOI{\tempurl}


\bibitem[\protect\citeauthoryear{Javed, Ahmed, Alazab, Jalil, Kifayat, and
  Gadekallu}{Javed et~al\mbox{.}}{2022}]%
        {Javed2022ADirections}
\bibfield{author}{\bibinfo{person}{Abdul~Rehman Javed}, \bibinfo{person}{Waqas
  Ahmed}, \bibinfo{person}{Mamoun Alazab}, \bibinfo{person}{Zunera Jalil},
  \bibinfo{person}{Kashif Kifayat}, {and} \bibinfo{person}{Thippa~Reddy
  Gadekallu}.} \bibinfo{year}{2022}\natexlab{}.
\newblock \showarticletitle{{A Comprehensive Survey on Computer Forensics:
  State-of-the-Art, Tools, Techniques, Challenges, and Future Directions}}.
\newblock \bibinfo{journal}{\emph{IEEE Access}}  \bibinfo{volume}{10}
  (\bibinfo{year}{2022}), \bibinfo{pages}{11065--11089}.
\newblock
\showISSN{2169-3536}
\urldef\tempurl%
\url{https://doi.org/10.1109/ACCESS.2022.3142508}
\showDOI{\tempurl}


\bibitem[\protect\citeauthoryear{Jhaver, Appling, Gilbert, and Bruckman}{Jhaver
  et~al\mbox{.}}{2019a}]%
        {Jhaver2019DidReddit}
\bibfield{author}{\bibinfo{person}{Shagun Jhaver},
  \bibinfo{person}{Darren~Scott Appling}, \bibinfo{person}{Eric Gilbert}, {and}
  \bibinfo{person}{Amy Bruckman}.} \bibinfo{year}{2019}\natexlab{a}.
\newblock \showarticletitle{{“Did you suspect the post would be removed?”:
  Understanding user reactions to content removals on reddit}}.
\newblock \bibinfo{journal}{\emph{Proceedings of the ACM on Human-Computer
  Interaction}} \bibinfo{volume}{3}, \bibinfo{number}{CSCW} (\bibinfo{date}{11}
  \bibinfo{year}{2019}), \bibinfo{pages}{1--33}.
\newblock
\showISSN{25730142}
\urldef\tempurl%
\url{https://doi.org/10.1145/3359294}
\showDOI{\tempurl}


\bibitem[\protect\citeauthoryear{Jhaver, Birman, Gilbert, and Bruckman}{Jhaver
  et~al\mbox{.}}{2019b}]%
        {Jhaver2019Human-machineAutomoderator}
\bibfield{author}{\bibinfo{person}{Shagun Jhaver}, \bibinfo{person}{Iris
  Birman}, \bibinfo{person}{Eric Gilbert}, {and} \bibinfo{person}{Amy
  Bruckman}.} \bibinfo{year}{2019}\natexlab{b}.
\newblock \showarticletitle{{Human-machine collaboration for content
  regulation: The case of reddit automoderator}}.
\newblock \bibinfo{journal}{\emph{ACM Transactions on Computer-Human
  Interaction}} \bibinfo{volume}{26}, \bibinfo{number}{5}
  (\bibinfo{year}{2019}), \bibinfo{pages}{35}.
\newblock
\showISBNx{10.1145/3338243}
\showISSN{15577325}
\urldef\tempurl%
\url{https://doi.org/10.1145/3338243}
\showDOI{\tempurl}


\bibitem[\protect\citeauthoryear{Jhaver, Boylston, Yang, and Bruckman}{Jhaver
  et~al\mbox{.}}{2021}]%
        {Jhaver2021EvaluatingTwitter}
\bibfield{author}{\bibinfo{person}{Shagun Jhaver}, \bibinfo{person}{Christian
  Boylston}, \bibinfo{person}{DIyi Yang}, {and} \bibinfo{person}{Amy
  Bruckman}.} \bibinfo{year}{2021}\natexlab{}.
\newblock \showarticletitle{{Evaluating the Effectiveness of Deplatforming as a
  Moderation Strategy on Twitter}}.
\newblock \bibinfo{journal}{\emph{Proceedings of the ACM on Human-Computer
  Interaction}} \bibinfo{volume}{5}, \bibinfo{number}{CSCW2}
  (\bibinfo{date}{10} \bibinfo{year}{2021}), \bibinfo{pages}{1--30}.
\newblock
\showISSN{2573-0142}
\urldef\tempurl%
\url{https://doi.org/10.1145/3479525}
\showDOI{\tempurl}


\bibitem[\protect\citeauthoryear{Jhaver, Chen, Knauss, and Zhang}{Jhaver
  et~al\mbox{.}}{2022}]%
        {Jhaver2022DesigningModeration}
\bibfield{author}{\bibinfo{person}{Shagun Jhaver}, \bibinfo{person}{Quan~Ze
  Chen}, \bibinfo{person}{Detlef Knauss}, {and} \bibinfo{person}{Amy~X.
  Zhang}.} \bibinfo{year}{2022}\natexlab{}.
\newblock \showarticletitle{{Designing Word Filter Tools for Creator-led
  Comment Moderation}}. In \bibinfo{booktitle}{\emph{CHI Conference on Human
  Factors in Computing Systems}}. \bibinfo{publisher}{ACM},
  \bibinfo{address}{New York, NY, USA}, \bibinfo{pages}{1--21}.
\newblock
\showISBNx{9781450391573}
\urldef\tempurl%
\url{https://doi.org/10.1145/3491102.3517505}
\showDOI{\tempurl}


\bibitem[\protect\citeauthoryear{Jiang, Kiene, Middler, Brubaker, and
  Fiesler}{Jiang et~al\mbox{.}}{2019}]%
        {Jiang2019ModerationDiscord}
\bibfield{author}{\bibinfo{person}{Jialun Jiang}, \bibinfo{person}{Charles
  Kiene}, \bibinfo{person}{Skyler Middler}, \bibinfo{person}{Jed~R. Brubaker},
  {and} \bibinfo{person}{Casey Fiesler}.} \bibinfo{year}{2019}\natexlab{}.
\newblock \showarticletitle{{Moderation challenges in voice-based online
  communities on discord}}.
\newblock \bibinfo{journal}{\emph{Proceedings of the ACM on Human-Computer
  Interaction}} \bibinfo{volume}{3}, \bibinfo{number}{CSCW} (\bibinfo{date}{11}
  \bibinfo{year}{2019}).
\newblock
\showISSN{25730142}
\urldef\tempurl%
\url{https://doi.org/10.1145/3359157}
\showDOI{\tempurl}


\bibitem[\protect\citeauthoryear{Jiang, Nie, Brubaker, and Fiesler}{Jiang
  et~al\mbox{.}}{2022}]%
        {Jiang2022AModeration}
\bibfield{author}{\bibinfo{person}{Jialun~Aaron Jiang}, \bibinfo{person}{Peipei
  Nie}, \bibinfo{person}{Jed~R Brubaker}, {and} \bibinfo{person}{Casey
  Fiesler}.} \bibinfo{year}{2022}\natexlab{}.
\newblock \showarticletitle{{A Trade-off-centered Framework of Content
  Moderation; A Trade-off-centered Framework of Content Moderation}}.
\newblock \bibinfo{journal}{\emph{ACM Trans. Comput.-Hum. Interact.}}
  (\bibinfo{year}{2022}), \bibinfo{pages}{1--34}.
\newblock
\urldef\tempurl%
\url{https://doi.org/10.1145/3534929}
\showDOI{\tempurl}


\bibitem[\protect\citeauthoryear{Jones, Dirckinck‐Holmfeld, and
  Lindstr{\"{o}}m}{Jones et~al\mbox{.}}{2006}]%
        {Jones2006ADecade}
\bibfield{author}{\bibinfo{person}{Chris Jones}, \bibinfo{person}{Lone
  Dirckinck‐Holmfeld}, {and} \bibinfo{person}{Berner Lindstr{\"{o}}m}.}
  \bibinfo{year}{2006}\natexlab{}.
\newblock \showarticletitle{{A relational, indirect, meso-level approach to
  CSCL design in the next decade}}.
\newblock \bibinfo{journal}{\emph{International Journal of Computer-Supported
  Collaborative Learning}} \bibinfo{volume}{1}, \bibinfo{number}{1}
  (\bibinfo{date}{3} \bibinfo{year}{2006}), \bibinfo{pages}{35--56}.
\newblock
\showISSN{1556-1607}
\urldef\tempurl%
\url{https://doi.org/10.1007/s11412-006-6841-7}
\showDOI{\tempurl}


\bibitem[\protect\citeauthoryear{Keyton and Menzie}{Keyton and Menzie}{2007}]%
        {Keyton2007SexuallyConversation}
\bibfield{author}{\bibinfo{person}{Joann Keyton} {and} \bibinfo{person}{Kathy
  Menzie}.} \bibinfo{year}{2007}\natexlab{}.
\newblock \showarticletitle{{Sexually Harassing Messages: Decoding Workplace
  Conversation}}.
\newblock \bibinfo{journal}{\emph{Communication Studies}} \bibinfo{volume}{58},
  \bibinfo{number}{1} (\bibinfo{date}{2} \bibinfo{year}{2007}),
  \bibinfo{pages}{87--103}.
\newblock
\showISSN{1051-0974}
\urldef\tempurl%
\url{https://doi.org/10.1080/10510970601168756}
\showDOI{\tempurl}


\bibitem[\protect\citeauthoryear{Kim, Oh, Cho, Shin, Suh, and Lee}{Kim
  et~al\mbox{.}}{2021}]%
        {Kim2021TrkicAlgorithms}
\bibfield{author}{\bibinfo{person}{Soomin Kim}, \bibinfo{person}{Changhoon Oh},
  \bibinfo{person}{Won~Ik Cho}, \bibinfo{person}{Donghoon Shin},
  \bibinfo{person}{Bongwon Suh}, {and} \bibinfo{person}{Joonhwan Lee}.}
  \bibinfo{year}{2021}\natexlab{}.
\newblock \showarticletitle{{Trkic G00gle: Why and How Users Game Translation
  Algorithms}}.
\newblock \bibinfo{journal}{\emph{Proceedings of the ACM on Human-Computer
  Interaction}} \bibinfo{volume}{5}, \bibinfo{number}{CSCW2}
  (\bibinfo{date}{10} \bibinfo{year}{2021}), \bibinfo{pages}{1--24}.
\newblock
\showISSN{2573-0142}
\urldef\tempurl%
\url{https://doi.org/10.1145/3476085}
\showDOI{\tempurl}


\bibitem[\protect\citeauthoryear{Kou}{Kou}{2021}]%
        {Kou2021PunishmentCommunity}
\bibfield{author}{\bibinfo{person}{Yubo Kou}.} \bibinfo{year}{2021}\natexlab{}.
\newblock \showarticletitle{{Punishment and Its Discontents: An Analysis of
  Permanent Ban in an Online Game Community}}.
\newblock \bibinfo{journal}{\emph{Proceedings of the ACM on Human-Computer
  Interaction}} \bibinfo{volume}{5}, \bibinfo{number}{CSCW2}
  (\bibinfo{date}{10} \bibinfo{year}{2021}), \bibinfo{pages}{1--21}.
\newblock
\showISSN{2573-0142}
\urldef\tempurl%
\url{https://doi.org/10.1145/3476075}
\showDOI{\tempurl}


\bibitem[\protect\citeauthoryear{Kumar, Hamilton, Leskovec, and Jurafsky}{Kumar
  et~al\mbox{.}}{2018}]%
        {Kumar2018CommunityWeb}
\bibfield{author}{\bibinfo{person}{Srijan Kumar}, \bibinfo{person}{William~L.
  Hamilton}, \bibinfo{person}{Jure Leskovec}, {and} \bibinfo{person}{Dan
  Jurafsky}.} \bibinfo{year}{2018}\natexlab{}.
\newblock \showarticletitle{{Community Interaction and Conflict on the Web}}.
  In \bibinfo{booktitle}{\emph{Proceedings of the 2018 World Wide Web
  Conference on World Wide Web}}. \bibinfo{publisher}{ACM Press},
  \bibinfo{address}{New York, New York, USA}, \bibinfo{pages}{933--943}.
\newblock
\showISBNx{9781450356398}
\urldef\tempurl%
\url{https://doi.org/10.1145/3178876.3186141}
\showDOI{\tempurl}


\bibitem[\protect\citeauthoryear{Kuznekoff and Rose}{Kuznekoff and
  Rose}{2013}]%
        {Kuznekoff2013CommunicationCues}
\bibfield{author}{\bibinfo{person}{Jeffrey~H. Kuznekoff} {and}
  \bibinfo{person}{Lindsey~M. Rose}.} \bibinfo{year}{2013}\natexlab{}.
\newblock \showarticletitle{{Communication in multiplayer gaming: Examining
  player responses to gender cues}}.
\newblock \bibinfo{journal}{\emph{New Media and Society}} \bibinfo{volume}{15},
  \bibinfo{number}{4} (\bibinfo{date}{9} \bibinfo{year}{2013}),
  \bibinfo{pages}{541--556}.
\newblock
\showISSN{14614448}
\urldef\tempurl%
\url{https://doi.org/10.1177/1461444812458271}
\showDOI{\tempurl}


\bibitem[\protect\citeauthoryear{Lai, Carton, Bhatnagar, Liao, Zhang, and
  Tan}{Lai et~al\mbox{.}}{2022}]%
        {Lai2022Human-AIModeration}
\bibfield{author}{\bibinfo{person}{Vivian Lai}, \bibinfo{person}{Samuel
  Carton}, \bibinfo{person}{Rajat Bhatnagar}, \bibinfo{person}{Q.~Vera Liao},
  \bibinfo{person}{Yunfeng Zhang}, {and} \bibinfo{person}{Chenhao Tan}.}
  \bibinfo{year}{2022}\natexlab{}.
\newblock \showarticletitle{{Human-AI Collaboration via Conditional Delegation:
  A Case Study of Content Moderation}}. In \bibinfo{booktitle}{\emph{CHI
  Conference on Human Factors in Computing Systems}}. \bibinfo{publisher}{ACM},
  \bibinfo{address}{New York, NY, USA}, \bibinfo{pages}{1--18}.
\newblock
\showISBNx{9781450391573}
\urldef\tempurl%
\url{https://doi.org/10.1145/3491102.3501999}
\showDOI{\tempurl}


\bibitem[\protect\citeauthoryear{Lambert, Rajagopal, and
  Chandrasekharan}{Lambert et~al\mbox{.}}{2022}]%
        {Lambert2022ConversationalEvents}
\bibfield{author}{\bibinfo{person}{Charlotte Lambert}, \bibinfo{person}{Ananya
  Rajagopal}, {and} \bibinfo{person}{Eshwar Chandrasekharan}.}
  \bibinfo{year}{2022}\natexlab{}.
\newblock \showarticletitle{{Conversational Resilience: Quantifying and
  Predicting Conversational Outcomes Following Adverse Events}}.
\newblock \bibinfo{journal}{\emph{Proceedings of the International AAAI
  Conference on Web and Social Media}} \bibinfo{volume}{16},
  \bibinfo{number}{1} (\bibinfo{year}{2022}), \bibinfo{pages}{548--559}.
\newblock
\urldef\tempurl%
\url{https://ojs.aaai.org/index.php/ICWSM/article/view/19314}
\showURL{%
\tempurl}


\bibitem[\protect\citeauthoryear{Lampe and Resnick}{Lampe and Resnick}{2004}]%
        {Lampe2004SlashdotSpace}
\bibfield{author}{\bibinfo{person}{Cliff Lampe} {and} \bibinfo{person}{Paul
  Resnick}.} \bibinfo{year}{2004}\natexlab{}.
\newblock \showarticletitle{{Slash(dot) and Burn: Distributed Moderation in a
  Large Online Conversation Space}}. In \bibinfo{booktitle}{\emph{Proceedings
  of the ACM Conference on Human Factors in Computing Systems}}.
  \bibinfo{pages}{543--550}.
\newblock
\showISBNx{1581137028}
\urldef\tempurl%
\url{https://doi.org/10.1145/985692.985761}
\showDOI{\tempurl}


\bibitem[\protect\citeauthoryear{Lee}{Lee}{2022}]%
        {Lee2022DontRaids}
\bibfield{author}{\bibinfo{person}{Alexander Lee}.}
  \bibinfo{year}{2022}\natexlab{}.
\newblock \bibinfo{title}{{‘Don’t let it bother you, just continue
  streaming’: Confessions of a Twitch streamer who received ‘hate
  raids’}}.
\newblock
\newblock
\urldef\tempurl%
\url{https://digiday.com/marketing/dont-let-it-bother-you-just-continue-streaming-confessions-of-a-twitch-streamer-and-victim-of-online-hate-raids/}
\showURL{%
\tempurl}


\bibitem[\protect\citeauthoryear{Lessel, Vielhauer, and Kr{\"{u}}ger}{Lessel
  et~al\mbox{.}}{2017}]%
        {Lessel2017ExpandingStudy}
\bibfield{author}{\bibinfo{person}{Pascal Lessel}, \bibinfo{person}{Alexander
  Vielhauer}, {and} \bibinfo{person}{Antonio Kr{\"{u}}ger}.}
  \bibinfo{year}{2017}\natexlab{}.
\newblock \showarticletitle{{Expanding video game live-streams with enhanced
  communication channels: A case study}}. In
  \bibinfo{booktitle}{\emph{Conference on Human Factors in Computing Systems -
  Proceedings}}. \bibinfo{publisher}{Association for Computing Machinery},
  \bibinfo{pages}{1571--1576}.
\newblock
\showISBNx{9781450346559}
\urldef\tempurl%
\url{https://doi.org/10.1145/3025453.3025708}
\showDOI{\tempurl}


\bibitem[\protect\citeauthoryear{Li, Hecht, and Chancellor}{Li
  et~al\mbox{.}}{2022}]%
        {Li2022MeasuringWork}
\bibfield{author}{\bibinfo{person}{Hanlin Li}, \bibinfo{person}{Brent Hecht},
  {and} \bibinfo{person}{Stevie Chancellor}.} \bibinfo{year}{2022}\natexlab{}.
\newblock \showarticletitle{{Measuring the Monetary Value of Online Volunteer
  Work}}. In \bibinfo{booktitle}{\emph{Proceedings of the International AAAI
  Conference on Web and Social Media}}. \bibinfo{pages}{596--606}.
\newblock
\urldef\tempurl%
\url{https://ojs.aaai.org/index.php/ICWSM/article/view/19318}
\showURL{%
\tempurl}


\bibitem[\protect\citeauthoryear{Li, Vincent, Tsai, Kaye, and Hecht}{Li
  et~al\mbox{.}}{2019}]%
        {Li2019HowProtest}
\bibfield{author}{\bibinfo{person}{Hanlin Li}, \bibinfo{person}{Nicholas
  Vincent}, \bibinfo{person}{Janice Tsai}, \bibinfo{person}{Jofish Kaye}, {and}
  \bibinfo{person}{Brent Hecht}.} \bibinfo{year}{2019}\natexlab{}.
\newblock \showarticletitle{{How Do People Change Their Technology Use in
  Protest?}}
\newblock \bibinfo{journal}{\emph{Proceedings of the ACM on Human-Computer
  Interaction}} \bibinfo{volume}{3}, \bibinfo{number}{CSCW} (\bibinfo{date}{11}
  \bibinfo{year}{2019}), \bibinfo{pages}{1--22}.
\newblock
\showISSN{2573-0142}
\urldef\tempurl%
\url{https://doi.org/10.1145/3359189}
\showDOI{\tempurl}


\bibitem[\protect\citeauthoryear{Li, Schulenberg, Freeman, and Acena}{Li
  et~al\mbox{.}}{2023}]%
        {Li2023WeReality}
\bibfield{author}{\bibinfo{person}{Lingyuan Li}, \bibinfo{person}{Kelsea
  Schulenberg}, \bibinfo{person}{Guo Freeman}, {and} \bibinfo{person}{Dane
  Acena}.} \bibinfo{year}{2023}\natexlab{}.
\newblock \showarticletitle{{"We Cried on Each Other's Shoulders": How LGBTQ+
  Individuals Experience Social Support in Social Virtual Reality}}. In
  \bibinfo{booktitle}{\emph{Proceedings of the 2023 CHI Conference on Human
  Factors in Computing Systems (CHI ’23)}}, Vol.~\bibinfo{volume}{16}.
  \bibinfo{publisher}{ACM}, \bibinfo{pages}{1--16}.
\newblock
\showISBNx{9781450394215}
\urldef\tempurl%
\url{https://doi.org/10.1145/3544548.3581530}
\showDOI{\tempurl}


\bibitem[\protect\citeauthoryear{Lindsay, Booth, Messing, and Thaller}{Lindsay
  et~al\mbox{.}}{2016}]%
        {Lindsay2016ExperiencesAdults}
\bibfield{author}{\bibinfo{person}{Megan Lindsay}, \bibinfo{person}{Jaime~M.
  Booth}, \bibinfo{person}{Jill~T. Messing}, {and} \bibinfo{person}{Jonel
  Thaller}.} \bibinfo{year}{2016}\natexlab{}.
\newblock \showarticletitle{{Experiences of Online Harassment Among Emerging
  Adults}}.
\newblock \bibinfo{journal}{\emph{Journal of Interpersonal Violence}}
  \bibinfo{volume}{31}, \bibinfo{number}{19} (\bibinfo{date}{11}
  \bibinfo{year}{2016}), \bibinfo{pages}{3174--3195}.
\newblock
\showISSN{0886-2605}
\urldef\tempurl%
\url{https://doi.org/10.1177/0886260515584344}
\showDOI{\tempurl}


\bibitem[\protect\citeauthoryear{Lopez and Freeman}{Lopez and Freeman}{2022}]%
        {Lopez2022ToStreaming}
\bibfield{author}{\bibinfo{person}{Jeremy Lopez} {and} \bibinfo{person}{Guo
  Freeman}.} \bibinfo{year}{2022}\natexlab{}.
\newblock \showarticletitle{{To Tag or Not To Tag: The Interplay of the Twitch
  Tag System and LGBTQIA+ Visibility in Live Streaming}}.
\newblock \bibinfo{journal}{\emph{Proceedings of the 55th Hawaii International
  Conference on System Sciences}} (\bibinfo{date}{1} \bibinfo{year}{2022}).
\newblock
\showISBNx{978-0-9981331-5-7}
\urldef\tempurl%
\url{https://doi.org/10.24251/HICSS.2022.413}
\showDOI{\tempurl}


\bibitem[\protect\citeauthoryear{Luo, Hsu, Park, and Hancock}{Luo
  et~al\mbox{.}}{2020}]%
        {Luo2020EmotionalEvents}
\bibfield{author}{\bibinfo{person}{Mufan Luo}, \bibinfo{person}{Tiffany~W.
  Hsu}, \bibinfo{person}{Joon~Sung Park}, {and} \bibinfo{person}{Jeffrey~T.
  Hancock}.} \bibinfo{year}{2020}\natexlab{}.
\newblock \showarticletitle{{Emotional Amplification During Live-Streaming:
  Evidence from Comments During and After News Events}}.
\newblock \bibinfo{journal}{\emph{Proceedings of the ACM on Human-Computer
  Interaction}} \bibinfo{volume}{4}, \bibinfo{number}{CSCW1} (\bibinfo{date}{5}
  \bibinfo{year}{2020}), \bibinfo{pages}{1--19}.
\newblock
\showISSN{2573-0142}
\urldef\tempurl%
\url{https://doi.org/10.1145/3392853}
\showDOI{\tempurl}


\bibitem[\protect\citeauthoryear{Mariconti, Suarez-Tangil, Blackburn,
  De~Cristofaro, Kourtellis, Leontiadis, Serrano, and Stringhini}{Mariconti
  et~al\mbox{.}}{2019}]%
        {Mariconti2019YouAttacks}
\bibfield{author}{\bibinfo{person}{Enrico Mariconti},
  \bibinfo{person}{Guillermo Suarez-Tangil}, \bibinfo{person}{Jeremy
  Blackburn}, \bibinfo{person}{Emiliano De~Cristofaro},
  \bibinfo{person}{Nicolas Kourtellis}, \bibinfo{person}{Ilias Leontiadis},
  \bibinfo{person}{Jordi~Luque Serrano}, {and} \bibinfo{person}{Gianluca
  Stringhini}.} \bibinfo{year}{2019}\natexlab{}.
\newblock \showarticletitle{{“You Know What to Do”: Proactive Detection of
  YouTube Videos Targeted by Coordinated Hate Attacks}}.
\newblock \bibinfo{journal}{\emph{Proceedings of the ACM on Human-Computer
  Interaction}} \bibinfo{volume}{3}, \bibinfo{number}{CSCW} (\bibinfo{date}{11}
  \bibinfo{year}{2019}), \bibinfo{pages}{1--21}.
\newblock
\showISSN{2573-0142}
\urldef\tempurl%
\url{https://doi.org/10.1145/3359309}
\showDOI{\tempurl}


\bibitem[\protect\citeauthoryear{Massanari}{Massanari}{2017}]%
        {Massanari2017GamergateTechnocultures}
\bibfield{author}{\bibinfo{person}{Adrienne Massanari}.}
  \bibinfo{year}{2017}\natexlab{}.
\newblock \showarticletitle{{{\#}Gamergate and The Fappening: How Reddit’s
  algorithm, governance, and culture support toxic technocultures}}.
\newblock \bibinfo{journal}{\emph{New Media and Society}} \bibinfo{volume}{19},
  \bibinfo{number}{3} (\bibinfo{year}{2017}), \bibinfo{pages}{329--346}.
\newblock
\showISSN{14617315}
\urldef\tempurl%
\url{https://doi.org/10.1177/1461444815608807}
\showDOI{\tempurl}


\bibitem[\protect\citeauthoryear{Mathew, Saha, Tharad, Rajgaria, Singhania,
  Maity, Goyal, and Mukherjee}{Mathew et~al\mbox{.}}{2019}]%
        {Mathew2019ThouSpeech}
\bibfield{author}{\bibinfo{person}{Binny Mathew}, \bibinfo{person}{Punyajoy
  Saha}, \bibinfo{person}{Hardik Tharad}, \bibinfo{person}{Subham Rajgaria},
  \bibinfo{person}{Prajwal Singhania}, \bibinfo{person}{Suman~Kalyan Maity},
  \bibinfo{person}{Pawan Goyal}, {and} \bibinfo{person}{Animesh Mukherjee}.}
  \bibinfo{year}{2019}\natexlab{}.
\newblock \showarticletitle{{Thou shalt not hate: Countering online hate
  speech}}. In \bibinfo{booktitle}{\emph{Proceedings of the 13th International
  Conference on Web and Social Media, ICWSM 2019}}. \bibinfo{pages}{369--380}.
\newblock
\urldef\tempurl%
\url{https://doi.org/10.13140/RG.2.2.31128.85765}
\showDOI{\tempurl}


\bibitem[\protect\citeauthoryear{McInnis, Ajmani, Sun, Hou, Zeng, and
  Dow}{McInnis et~al\mbox{.}}{2021}]%
        {McInnis2021ReportingWebsite}
\bibfield{author}{\bibinfo{person}{Brian McInnis}, \bibinfo{person}{Leah
  Ajmani}, \bibinfo{person}{Lu Sun}, \bibinfo{person}{Yiwen Hou},
  \bibinfo{person}{Ziwen Zeng}, {and} \bibinfo{person}{Steven~P. Dow}.}
  \bibinfo{year}{2021}\natexlab{}.
\newblock \showarticletitle{{Reporting the Community Beat: Practices for
  Moderating Online Discussion at a News Website}}.
\newblock \bibinfo{journal}{\emph{Proceedings of the ACM on Human-Computer
  Interaction}} \bibinfo{volume}{5}, \bibinfo{number}{CSCW2}
  (\bibinfo{date}{10} \bibinfo{year}{2021}), \bibinfo{pages}{1--25}.
\newblock
\showISSN{2573-0142}
\urldef\tempurl%
\url{https://doi.org/10.1145/3476074}
\showDOI{\tempurl}


\bibitem[\protect\citeauthoryear{McLean and Griffiths}{McLean and
  Griffiths}{2019}]%
        {McLean2019FemaleStudy}
\bibfield{author}{\bibinfo{person}{Lavinia McLean} {and}
  \bibinfo{person}{Mark~D. Griffiths}.} \bibinfo{year}{2019}\natexlab{}.
\newblock \showarticletitle{{Female Gamers’ Experience of Online Harassment
  and Social Support in Online Gaming: A Qualitative Study}}.
\newblock \bibinfo{journal}{\emph{International Journal of Mental Health and
  Addiction}} \bibinfo{volume}{17}, \bibinfo{number}{4} (\bibinfo{date}{8}
  \bibinfo{year}{2019}), \bibinfo{pages}{970--994}.
\newblock
\showISSN{15571882}
\urldef\tempurl%
\url{https://doi.org/10.1007/S11469-018-9962-0/FIGURES/1}
\showDOI{\tempurl}


\bibitem[\protect\citeauthoryear{Meredith}{Meredith}{2017}]%
        {Meredith2017AnalysingAnalysis}
\bibfield{author}{\bibinfo{person}{Joanne Meredith}.}
  \bibinfo{year}{2017}\natexlab{}.
\newblock \showarticletitle{{Analysing technological affordances of online
  interactions using conversation analysis}}.
\newblock \bibinfo{journal}{\emph{Journal of Pragmatics}}
  \bibinfo{volume}{115} (\bibinfo{date}{7} \bibinfo{year}{2017}),
  \bibinfo{pages}{42--55}.
\newblock
\showISSN{03782166}
\urldef\tempurl%
\url{https://doi.org/10.1016/j.pragma.2017.03.001}
\showDOI{\tempurl}


\bibitem[\protect\citeauthoryear{Mir{\'{o}}-Llinares and
  Moneva}{Mir{\'{o}}-Llinares and Moneva}{2019}]%
        {Miro-Llinares2019WhatDrop}
\bibfield{author}{\bibinfo{person}{Fernando Mir{\'{o}}-Llinares} {and}
  \bibinfo{person}{Asier Moneva}.} \bibinfo{year}{2019}\natexlab{}.
\newblock \showarticletitle{{What about cyberspace (and cybercrime alongside
  it)? A reply to Farrell and Birks “Did cybercrime cause the crime
  drop?”}}.
\newblock \bibinfo{journal}{\emph{Crime Science}} \bibinfo{volume}{8},
  \bibinfo{number}{1} (\bibinfo{date}{12} \bibinfo{year}{2019}),
  \bibinfo{pages}{12}.
\newblock
\showISSN{2193-7680}
\urldef\tempurl%
\url{https://doi.org/10.1186/s40163-019-0107-y}
\showDOI{\tempurl}


\bibitem[\protect\citeauthoryear{Mitchell, Finkelhor, Jones, and
  Wolak}{Mitchell et~al\mbox{.}}{2010}]%
        {Mitchell2010UseUtilization.}
\bibfield{author}{\bibinfo{person}{Kimberly~J Mitchell}, \bibinfo{person}{David
  Finkelhor}, \bibinfo{person}{Lisa~M Jones}, {and} \bibinfo{person}{Janis
  Wolak}.} \bibinfo{year}{2010}\natexlab{}.
\newblock \showarticletitle{{Use of social networking sites in online sex
  crimes against minors: an examination of national incidence and means of
  utilization.}}
\newblock \bibinfo{journal}{\emph{The Journal of adolescent health : official
  publication of the Society for Adolescent Medicine}} \bibinfo{volume}{47},
  \bibinfo{number}{2} (\bibinfo{date}{8} \bibinfo{year}{2010}),
  \bibinfo{pages}{183--90}.
\newblock
\showISSN{1879-1972}
\urldef\tempurl%
\url{https://doi.org/10.1016/j.jadohealth.2010.01.007}
\showDOI{\tempurl}


\bibitem[\protect\citeauthoryear{Nightingale}{Nightingale}{2022}]%
        {Nightingale2022TwitchHarassment}
\bibfield{author}{\bibinfo{person}{Ed Nightingale}.}
  \bibinfo{year}{2022}\natexlab{}.
\newblock \bibinfo{title}{{Twitch streamers positive about new raid feature in
  response to harassment }}.
\newblock
\newblock
\urldef\tempurl%
\url{https://www.eurogamer.net/twitch-streamers-positive-about-new-raid-feature-in-response-to-harassment}
\showURL{%
\tempurl}


\bibitem[\protect\citeauthoryear{Nilizadeh, Labr{\`{e}}che, Sedighian, Zand,
  Fernandez, Kruegel, Stringhini, and Vigna}{Nilizadeh et~al\mbox{.}}{2017}]%
        {Nilizadeh2017POISED:Paths}
\bibfield{author}{\bibinfo{person}{Shirin Nilizadeh},
  \bibinfo{person}{François Labr{\`{e}}che}, \bibinfo{person}{Alireza
  Sedighian}, \bibinfo{person}{Ali Zand}, \bibinfo{person}{José Fernandez},
  \bibinfo{person}{Christopher Kruegel}, \bibinfo{person}{Gianluca Stringhini},
  {and} \bibinfo{person}{Giovanni Vigna}.} \bibinfo{year}{2017}\natexlab{}.
\newblock \showarticletitle{{POISED: Spotting twitter spam off the beaten
  paths}}. In \bibinfo{booktitle}{\emph{Proceedings of the ACM Conference on
  Computer and Communications Security}}. \bibinfo{publisher}{ACM},
  \bibinfo{address}{New York, NY, USA}, \bibinfo{pages}{1159--1174}.
\newblock
\showISBNx{9781450349468}
\showISSN{15437221}
\urldef\tempurl%
\url{https://doi.org/10.1145/3133956.3134055}
\showDOI{\tempurl}


\bibitem[\protect\citeauthoryear{Norman}{Norman}{1988}]%
        {Norman1988TheThings}
\bibfield{author}{\bibinfo{person}{Donald~A. Norman}.}
  \bibinfo{year}{1988}\natexlab{}.
\newblock \bibinfo{booktitle}{\emph{{The Psychology of Everyday Things}}}.
\newblock \bibinfo{publisher}{Basic Books}, \bibinfo{address}{New York}.
\newblock
\urldef\tempurl%
\url{https://psycnet.apa.org/record/1988-97561-000}
\showURL{%
\tempurl}


\bibitem[\protect\citeauthoryear{Page, Capener, Cullen, Wang, Garfield, and
  J.~Wisniewski}{Page et~al\mbox{.}}{2022}]%
        {Page2022PerceivingMedia}
\bibfield{author}{\bibinfo{person}{Xinru Page}, \bibinfo{person}{Andrew
  Capener}, \bibinfo{person}{Spring Cullen}, \bibinfo{person}{Tao Wang},
  \bibinfo{person}{Monica Garfield}, {and} \bibinfo{person}{Pamela
  J.~Wisniewski}.} \bibinfo{year}{2022}\natexlab{}.
\newblock \showarticletitle{{Perceiving Affordances Differently: The Unintended
  Consequences When Young Autistic Adults Engage with Social Media}}. In
  \bibinfo{booktitle}{\emph{CHI Conference on Human Factors in Computing
  Systems}}. \bibinfo{publisher}{ACM}, \bibinfo{address}{New York, NY, USA},
  \bibinfo{pages}{1--21}.
\newblock
\showISBNx{9781450391573}
\urldef\tempurl%
\url{https://doi.org/10.1145/3491102.3517596}
\showDOI{\tempurl}


\bibitem[\protect\citeauthoryear{Park, Popowski, Cai, Morris, Liang, and
  Bernstein}{Park et~al\mbox{.}}{2022}]%
        {Park2022SocialSystems}
\bibfield{author}{\bibinfo{person}{Joon~Sung Park}, \bibinfo{person}{Lindsay
  Popowski}, \bibinfo{person}{Carrie~J Cai}, \bibinfo{person}{Meredith~Ringel
  Morris}, \bibinfo{person}{Percy Liang}, {and} \bibinfo{person}{Michael~S
  Bernstein}.} \bibinfo{year}{2022}\natexlab{}.
\newblock \showarticletitle{{Social Simulacra: Creating Populated Prototypes
  for Social Computing Systems}}. In \bibinfo{booktitle}{\emph{The 35th Annual
  ACM Symposium on User Interface Software and Technology (UIST '22)}}.
  \bibinfo{pages}{1--18}.
\newblock
\showISBNx{9781450393201}
\urldef\tempurl%
\url{https://doi.org/10.1145/3526113.3545616}
\showDOI{\tempurl}


\bibitem[\protect\citeauthoryear{Parrish}{Parrish}{2021a}]%
        {Parrish2021HowRaid}
\bibfield{author}{\bibinfo{person}{Ash Parrish}.}
  \bibinfo{year}{2021}\natexlab{a}.
\newblock \bibinfo{title}{{How to stop a hate raid}}.
\newblock
\newblock
\urldef\tempurl%
\url{https://www.theverge.com/22633874/how-to-stop-a-hate-raid-twitch-safety-tools}
\showURL{%
\tempurl}


\bibitem[\protect\citeauthoryear{Parrish}{Parrish}{2021b}]%
        {Parrish2021TwitchRaiders}
\bibfield{author}{\bibinfo{person}{Ash Parrish}.}
  \bibinfo{year}{2021}\natexlab{b}.
\newblock \bibinfo{title}{{Twitch sues two alleged ‘hate raiders’}}.
\newblock
\newblock
\urldef\tempurl%
\url{https://www.theverge.com/2021/9/10/22666953/twitch-sues-alleged-hate-raiders-harassment-streamers}
\showURL{%
\tempurl}


\bibitem[\protect\citeauthoryear{Quayle}{Quayle}{2016}]%
        {Quayle2016ResearchingVulnerabilities}
\bibfield{author}{\bibinfo{person}{Ethel Quayle}.}
  \bibinfo{year}{2016}\natexlab{}.
\newblock \bibinfo{booktitle}{\emph{{Researching online child sexual
  exploitation and abuse: Are there links between online and offline
  vulnerabilities?}}}
\newblock \bibinfo{type}{{T}echnical {R}eport}. \bibinfo{institution}{The
  University of Edinburgh}, \bibinfo{address}{UK}. \bibinfo{pages}{1--48}
  pages.
\newblock
\urldef\tempurl%
\url{www.globalkidsonline.net/sexual-}
\showURL{%
\tempurl}


\bibitem[\protect\citeauthoryear{Riddick and Shivener}{Riddick and
  Shivener}{2022}]%
        {Riddick2022AffectiveLivestream}
\bibfield{author}{\bibinfo{person}{Sarah Riddick} {and} \bibinfo{person}{Rich
  Shivener}.} \bibinfo{year}{2022}\natexlab{}.
\newblock \showarticletitle{{Affective Spamming on Twitch: Rhetorics of an
  Emote-Only Audience in a Presidential Inauguration Livestream}}.
\newblock \bibinfo{journal}{\emph{Computers and Composition}}
  \bibinfo{volume}{64} (\bibinfo{date}{6} \bibinfo{year}{2022}),
  \bibinfo{pages}{102711}.
\newblock
\showISSN{87554615}
\urldef\tempurl%
\url{https://doi.org/10.1016/j.compcom.2022.102711}
\showDOI{\tempurl}


\bibitem[\protect\citeauthoryear{Robey, Anderson, for, and 2013}{Robey
  et~al\mbox{.}}{2013}]%
        {Robey2013InformationOdyssey}
\bibfield{author}{\bibinfo{person}{D Robey}, \bibinfo{person}{C Anderson},
  \bibinfo{person}{B~Raymond Journal of the~Association for}, {and}
  \bibinfo{person}{undefined 2013}.} \bibinfo{year}{2013}\natexlab{}.
\newblock \showarticletitle{{Information technology, materiality, and
  organizational change: A professional odyssey}}.
\newblock \bibinfo{journal}{\emph{aisel.aisnet.org}} \bibinfo{volume}{14},
  \bibinfo{number}{7} (\bibinfo{year}{2013}), \bibinfo{pages}{379--398}.
\newblock
\urldef\tempurl%
\url{https://aisel.aisnet.org/jais/vol14/iss7/1/}
\showURL{%
\tempurl}


\bibitem[\protect\citeauthoryear{Scheuerman, Branham, and Hamidi}{Scheuerman
  et~al\mbox{.}}{2018}]%
        {Scheuerman2018SafePeople}
\bibfield{author}{\bibinfo{person}{Morgan~Klaus Scheuerman},
  \bibinfo{person}{Stacy~M. Branham}, {and} \bibinfo{person}{Foad Hamidi}.}
  \bibinfo{year}{2018}\natexlab{}.
\newblock \showarticletitle{{Safe spaces and safe places: Unpacking
  technology-mediated experiences of safety and harm with transgender people}}.
\newblock \bibinfo{journal}{\emph{Proceedings of the ACM on Human-Computer
  Interaction}} \bibinfo{volume}{2}, \bibinfo{number}{CSCW} (\bibinfo{date}{11}
  \bibinfo{year}{2018}), \bibinfo{pages}{1--27}.
\newblock
\showISSN{25730142}
\urldef\tempurl%
\url{https://doi.org/10.1145/3274424}
\showDOI{\tempurl}


\bibitem[\protect\citeauthoryear{Scheuerman, Jiang, Fiesler, and
  Brubaker}{Scheuerman et~al\mbox{.}}{2021}]%
        {Scheuerman2021AOnline}
\bibfield{author}{\bibinfo{person}{Morgan~Klaus Scheuerman},
  \bibinfo{person}{Jialun~Aaron Jiang}, \bibinfo{person}{Casey Fiesler}, {and}
  \bibinfo{person}{Jed~R. Brubaker}.} \bibinfo{year}{2021}\natexlab{}.
\newblock \showarticletitle{{A Framework of Severity for Harmful Content
  Online}}.
\newblock \bibinfo{journal}{\emph{Proceedings of the ACM on Human-Computer
  Interaction}} \bibinfo{volume}{5}, \bibinfo{number}{CSCW2}
  (\bibinfo{date}{10} \bibinfo{year}{2021}), \bibinfo{pages}{1--33}.
\newblock
\showISSN{2573-0142}
\urldef\tempurl%
\url{https://doi.org/10.1145/3479512}
\showDOI{\tempurl}


\bibitem[\protect\citeauthoryear{Schoenebeck and Blackwell}{Schoenebeck and
  Blackwell}{2021}]%
        {Schoenebeck2021ReimaginingRepair}
\bibfield{author}{\bibinfo{person}{Sarita Schoenebeck} {and}
  \bibinfo{person}{Lindsay Blackwell}.} \bibinfo{year}{2021}\natexlab{}.
\newblock \showarticletitle{{Reimagining Social Media Governance: Harm,
  Accountability, and Repair}}.
\newblock \bibinfo{journal}{\emph{SSRN Electronic Journal}} (\bibinfo{date}{7}
  \bibinfo{year}{2021}).
\newblock
\showISSN{1556-5068}
\urldef\tempurl%
\url{https://doi.org/10.2139/ssrn.3895779}
\showDOI{\tempurl}


\bibitem[\protect\citeauthoryear{Schulenberg, Li, Freeman, Zamanifard, and
  Mcneese}{Schulenberg et~al\mbox{.}}{2023}]%
        {Schulenberg2023TowardsReality}
\bibfield{author}{\bibinfo{person}{Kelsea Schulenberg},
  \bibinfo{person}{Lingyuan Li}, \bibinfo{person}{Guo Freeman},
  \bibinfo{person}{Samaneh Zamanifard}, {and} \bibinfo{person}{Nathan~J
  Mcneese}.} \bibinfo{year}{2023}\natexlab{}.
\newblock \showarticletitle{{Towards Leveraging AI-based Moderation to Address
  Emergent Harassment in Social Virtual Reality}}. In
  \bibinfo{booktitle}{\emph{Proceedings of the 2023 CHI Conference on Human
  Factors in Computing Systems (CHI '23)}}, Vol.~\bibinfo{volume}{1}.
  \bibinfo{publisher}{ACM}, \bibinfo{pages}{1--17}.
\newblock
\showISBNx{9781450394215}
\urldef\tempurl%
\url{https://doi.org/10.1145/3544548.3581090}
\showDOI{\tempurl}


\bibitem[\protect\citeauthoryear{Seering}{Seering}{2020}]%
        {Seering2020ReconsideringModeration}
\bibfield{author}{\bibinfo{person}{Joseph Seering}.}
  \bibinfo{year}{2020}\natexlab{}.
\newblock \showarticletitle{{Reconsidering Community Self-Moderation: The Role
  of Research in Supporting Community-Based Models for Online Content
  Moderation}}.
\newblock \bibinfo{journal}{\emph{Proceedings of the ACM on Human-Computer
  Interaction}} \bibinfo{volume}{4}, \bibinfo{number}{CSCW2}
  (\bibinfo{date}{10} \bibinfo{year}{2020}).
\newblock
\showISSN{25730142}
\urldef\tempurl%
\url{https://doi.org/10.1145/3415178}
\showDOI{\tempurl}


\bibitem[\protect\citeauthoryear{Seering, Fang, Damasco, Chen, Sun, and
  Kaufman}{Seering et~al\mbox{.}}{2019a}]%
        {Seering2019DesigningBehaviors}
\bibfield{author}{\bibinfo{person}{Joseph Seering}, \bibinfo{person}{Tianmi
  Fang}, \bibinfo{person}{Luca Damasco}, \bibinfo{person}{Mianhong~Cherie
  Chen}, \bibinfo{person}{Likang Sun}, {and} \bibinfo{person}{Geoff Kaufman}.}
  \bibinfo{year}{2019}\natexlab{a}.
\newblock \showarticletitle{{Designing User Interface Elements to Improve the
  Quality and Civility of Discourse in Online Commenting Behaviors}}. In
  \bibinfo{booktitle}{\emph{Proceedings of the 2019 CHI Conference on Human
  Factors in Computing Systems}}. \bibinfo{publisher}{ACM},
  \bibinfo{address}{New York, NY, USA}, \bibinfo{pages}{1--14}.
\newblock
\showISBNx{9781450359702}
\urldef\tempurl%
\url{https://doi.org/10.1145/3290605.3300836}
\showDOI{\tempurl}


\bibitem[\protect\citeauthoryear{Seering, Kraut, and Dabbish}{Seering
  et~al\mbox{.}}{2017}]%
        {Seering2017ShapingExample-Setting}
\bibfield{author}{\bibinfo{person}{Joseph Seering}, \bibinfo{person}{Robert
  Kraut}, {and} \bibinfo{person}{Laura Dabbish}.}
  \bibinfo{year}{2017}\natexlab{}.
\newblock \showarticletitle{{Shaping Pro and Anti-Social Behavior on Twitch
  Through Moderation and Example-Setting}}. In
  \bibinfo{booktitle}{\emph{Proceedings of the 2017 ACM Conference on Computer
  Supported Cooperative Work and Social Computing}}. \bibinfo{publisher}{ACM},
  \bibinfo{address}{New York, NY, USA}, \bibinfo{pages}{111--125}.
\newblock
\showISBNx{9781450343350}
\urldef\tempurl%
\url{https://doi.org/10.1145/2998181.2998277}
\showDOI{\tempurl}


\bibitem[\protect\citeauthoryear{Seering, Wang, Yoon, and Kaufman}{Seering
  et~al\mbox{.}}{2019b}]%
        {Seering2019ModeratorAlgorithms}
\bibfield{author}{\bibinfo{person}{Joseph Seering}, \bibinfo{person}{Tony
  Wang}, \bibinfo{person}{Jina Yoon}, {and} \bibinfo{person}{Geoff Kaufman}.}
  \bibinfo{year}{2019}\natexlab{b}.
\newblock \showarticletitle{{Moderator engagement and community development in
  the age of algorithms}}.
\newblock \bibinfo{journal}{\emph{New Media {\&} Society}}
  \bibinfo{volume}{21}, \bibinfo{number}{7} (\bibinfo{date}{7}
  \bibinfo{year}{2019}), \bibinfo{pages}{1417--1443}.
\newblock
\showISSN{1461-4448}
\urldef\tempurl%
\url{https://doi.org/10.1177/1461444818821316}
\showDOI{\tempurl}


\bibitem[\protect\citeauthoryear{Sheng and Kairam}{Sheng and Kairam}{2020}]%
        {Sheng2020FromTwitch}
\bibfield{author}{\bibinfo{person}{Jeff~T Sheng} {and}
  \bibinfo{person}{Sanjay~R Kairam}.} \bibinfo{year}{2020}\natexlab{}.
\newblock \showarticletitle{{From Virtual Strangers to IRL Friends:
  Relationship Development in Livestreaming Communities on Twitch}}.
\newblock \bibinfo{journal}{\emph{Proceedings of the ACM on Human-Computer
  Interaction}} \bibinfo{volume}{4}, \bibinfo{number}{CSCW2}
  (\bibinfo{year}{2020}), \bibinfo{pages}{34}.
\newblock
\showISSN{25730142}
\urldef\tempurl%
\url{https://doi.org/10.1145/3415165}
\showDOI{\tempurl}


\bibitem[\protect\citeauthoryear{Steiger, Bharucha, Venkatagiri, Riedl, and
  Lease}{Steiger et~al\mbox{.}}{2021}]%
        {Steiger2021TheSupport}
\bibfield{author}{\bibinfo{person}{Miriah Steiger}, \bibinfo{person}{Timir~J.
  Bharucha}, \bibinfo{person}{Sukrit Venkatagiri}, \bibinfo{person}{Martin~J.
  Riedl}, {and} \bibinfo{person}{Matthew Lease}.}
  \bibinfo{year}{2021}\natexlab{}.
\newblock \showarticletitle{{The Psychological Well-Being of Content
  Moderators: The Emotional Labor of Commercial Moderation and Avenues for
  Improving Support}}. In \bibinfo{booktitle}{\emph{Conference on Human Factors
  in Computing Systems - Proceedings}}. \bibinfo{publisher}{Association for
  Computing Machinery}, \bibinfo{pages}{1--14}.
\newblock
\showISBNx{9781450380966}
\urldef\tempurl%
\url{https://doi.org/10.1145/3411764.3445092}
\showDOI{\tempurl}


\bibitem[\protect\citeauthoryear{Stringhini, Mourlanne, Jacob, Egele, Kruegel,
  and Vigna}{Stringhini et~al\mbox{.}}{2015}]%
        {Stringhini2015Evilcohort:Services}
\bibfield{author}{\bibinfo{person}{Gianluca Stringhini},
  \bibinfo{person}{Pierre Mourlanne}, \bibinfo{person}{Gregoire Jacob},
  \bibinfo{person}{Manuel Egele}, \bibinfo{person}{Christopher Kruegel}, {and}
  \bibinfo{person}{Giovanni Vigna}.} \bibinfo{year}{2015}\natexlab{}.
\newblock \bibinfo{title}{{Evilcohort: Detecting communities of malicious
  accounts on online services}}.
\newblock , \bibinfo{numpages}{563--578}~pages.
\newblock
\showISBNx{9781931971232}
\urldef\tempurl%
\url{https://www.usenix.org/conference/usenixsecurity15/technical-sessions/presentation/stringhini}
\showURL{%
\tempurl}


\bibitem[\protect\citeauthoryear{Stuart}{Stuart}{2014}]%
        {Stuart2014ZoeLives}
\bibfield{author}{\bibinfo{person}{Keith Stuart}.}
  \bibinfo{year}{2014}\natexlab{}.
\newblock \bibinfo{title}{{Zoe Quinn: 'All Gamergate has done is ruin people's
  lives'}}.
\newblock
\newblock
\urldef\tempurl%
\url{https://www.theguardian.com/technology/2014/dec/03/zoe-quinn-gamergate-interview}
\showURL{%
\tempurl}


\bibitem[\protect\citeauthoryear{Sultana, Deb, Bhattacharjee, Hasan, Alam,
  Chakraborty, Roy, Ahmed, Moitra, Amin, Islam, and Ahmed}{Sultana
  et~al\mbox{.}}{2021}]%
        {Sultana2021Unmochon:Messenger}
\bibfield{author}{\bibinfo{person}{Sharifa Sultana}, \bibinfo{person}{Mitrasree
  Deb}, \bibinfo{person}{Ananya Bhattacharjee}, \bibinfo{person}{Shaid Hasan},
  \bibinfo{person}{S.M.Raihanul Alam}, \bibinfo{person}{Trishna Chakraborty},
  \bibinfo{person}{Prianka Roy}, \bibinfo{person}{Samira~Fairuz Ahmed},
  \bibinfo{person}{Aparna Moitra}, \bibinfo{person}{M.~Ashraful Amin},
  \bibinfo{person}{A.K.M.~Najmul Islam}, {and} \bibinfo{person}{Syed~Ishtiaque
  Ahmed}.} \bibinfo{year}{2021}\natexlab{}.
\newblock \showarticletitle{{‘Unmochon’: A Tool to Combat Online Sexual
  Harassment over Facebook Messenger}}. In
  \bibinfo{booktitle}{\emph{Proceedings of the 2021 CHI Conference on Human
  Factors in Computing Systems}}. \bibinfo{publisher}{ACM},
  \bibinfo{address}{New York, NY, USA}, \bibinfo{pages}{1--18}.
\newblock
\showISBNx{9781450380966}
\urldef\tempurl%
\url{https://doi.org/10.1145/3411764.3445154}
\showDOI{\tempurl}


\bibitem[\protect\citeauthoryear{Thach, Mayworm, Delmonaco, and Haimson}{Thach
  et~al\mbox{.}}{2022}]%
        {Thach2022InvisibleReddit}
\bibfield{author}{\bibinfo{person}{Hibby Thach}, \bibinfo{person}{Samuel
  Mayworm}, \bibinfo{person}{Daniel Delmonaco}, {and} \bibinfo{person}{Oliver
  Haimson}.} \bibinfo{year}{2022}\natexlab{}.
\newblock \showarticletitle{{(In)visible moderation: A digital ethnography of
  marginalized users and content moderation on Twitch and Reddit}}.
\newblock \bibinfo{journal}{\emph{New Media {\&} Society}} (\bibinfo{date}{7}
  \bibinfo{year}{2022}), \bibinfo{pages}{1--22}.
\newblock
\showISSN{1461-4448}
\urldef\tempurl%
\url{https://doi.org/10.1177/14614448221109804}
\showDOI{\tempurl}


\bibitem[\protect\citeauthoryear{Uttarapong, Cai, and Wohn}{Uttarapong
  et~al\mbox{.}}{2021}]%
        {Uttarapong2021HarassmentNegativity}
\bibfield{author}{\bibinfo{person}{Jirassaya Uttarapong}, \bibinfo{person}{Jie
  Cai}, {and} \bibinfo{person}{Donghee~Yvette Wohn}.}
  \bibinfo{year}{2021}\natexlab{}.
\newblock \showarticletitle{{Harassment Experiences of Women and LGBTQ Live
  Streamers and How They Handled Negativity}}. In \bibinfo{booktitle}{\emph{ACM
  International Conference on Interactive Media Experiences}}.
  \bibinfo{publisher}{ACM}, \bibinfo{address}{New York, NY, USA},
  \bibinfo{pages}{7--19}.
\newblock
\showISBNx{9781450383899}
\urldef\tempurl%
\url{https://doi.org/10.1145/3452918.3458794}
\showDOI{\tempurl}


\bibitem[\protect\citeauthoryear{Vitak, Chadha, Steiner, and Ashktorab}{Vitak
  et~al\mbox{.}}{2017}]%
        {Vitak2017IdentifyingHarassment}
\bibfield{author}{\bibinfo{person}{Jessica Vitak}, \bibinfo{person}{Kalyani
  Chadha}, \bibinfo{person}{Linda Steiner}, {and} \bibinfo{person}{Zahra
  Ashktorab}.} \bibinfo{year}{2017}\natexlab{}.
\newblock \showarticletitle{{Identifying women's experiences with and
  strategies for mitigating negative effects of online harassment}}. In
  \bibinfo{booktitle}{\emph{Proceedings of the ACM Conference on Computer
  Supported Cooperative Work}}. \bibinfo{pages}{1231--1245}.
\newblock
\showISBNx{9781450343350}
\urldef\tempurl%
\url{https://doi.org/10.1145/2998181.2998337}
\showDOI{\tempurl}


\bibitem[\protect\citeauthoryear{Walker and DeVito}{Walker and DeVito}{2020}]%
        {Walker2020MoreSpaces}
\bibfield{author}{\bibinfo{person}{Ashley~Marie Walker} {and}
  \bibinfo{person}{Michael~A. DeVito}.} \bibinfo{year}{2020}\natexlab{}.
\newblock \showarticletitle{{"'More gay' fits in better": Intracommunity Power
  Dynamics and Harms in Online LGBTQ+ Spaces}}. In
  \bibinfo{booktitle}{\emph{Proceedings of the 2020 CHI Conference on Human
  Factors in Computing Systems}}. \bibinfo{publisher}{ACM},
  \bibinfo{address}{New York, NY, USA}, \bibinfo{pages}{1--15}.
\newblock
\showISBNx{9781450367080}
\urldef\tempurl%
\url{https://doi.org/10.1145/3313831.3376497}
\showDOI{\tempurl}


\bibitem[\protect\citeauthoryear{Wohn}{Wohn}{2019}]%
        {Wohn2019VolunteerExperience}
\bibfield{author}{\bibinfo{person}{Donghee~Yvette Wohn}.}
  \bibinfo{year}{2019}\natexlab{}.
\newblock \showarticletitle{{Volunteer moderators in twitch micro communities:
  How they get involved, the roles they play, and the emotional labor they
  experience}}.
\newblock \bibinfo{journal}{\emph{Conference on Human Factors in Computing
  Systems - Proceedings}} (\bibinfo{date}{5} \bibinfo{year}{2019}).
\newblock
\showISBNx{9781450359702}
\urldef\tempurl%
\url{https://doi.org/10.1145/3290605.3300390}
\showDOI{\tempurl}


\bibitem[\protect\citeauthoryear{Wohn, Fiesler, Hemphill, De~Choudhury, and
  Matias}{Wohn et~al\mbox{.}}{2017}]%
        {Wohn2017HowMedia}
\bibfield{author}{\bibinfo{person}{Donghee~Yvette Wohn}, \bibinfo{person}{Casey
  Fiesler}, \bibinfo{person}{Libby Hemphill}, \bibinfo{person}{Munmun
  De~Choudhury}, {and} \bibinfo{person}{J.~Nathan Matias}.}
  \bibinfo{year}{2017}\natexlab{}.
\newblock \showarticletitle{{How to handle online risks? Discussing content
  curation and moderation in social media}}.
\newblock \bibinfo{journal}{\emph{Conference on Human Factors in Computing
  Systems - Proceedings}}  \bibinfo{volume}{Part F127655} (\bibinfo{date}{5}
  \bibinfo{year}{2017}), \bibinfo{pages}{1271--1276}.
\newblock
\showISBNx{9781450346566}
\urldef\tempurl%
\url{https://doi.org/10.1145/3027063.3051141}
\showDOI{\tempurl}


\bibitem[\protect\citeauthoryear{Wohn and Freeman}{Wohn and Freeman}{2020}]%
        {Wohn2020AudienceTwitch}
\bibfield{author}{\bibinfo{person}{Donghee~Yvette Wohn} {and}
  \bibinfo{person}{Guo Freeman}.} \bibinfo{year}{2020}\natexlab{}.
\newblock \showarticletitle{{Audience Management Practices of Live Streamers on
  Twitch}}.
\newblock \bibinfo{journal}{\emph{IMX 2020 - Proceedings of the 2020 ACM
  International Conference on Interactive Media Experiences}}
  (\bibinfo{date}{6} \bibinfo{year}{2020}), \bibinfo{pages}{106--116}.
\newblock
\showISBNx{9781450379762}
\urldef\tempurl%
\url{https://doi.org/10.1145/3391614.3393653}
\showDOI{\tempurl}


\bibitem[\protect\citeauthoryear{Woolley}{Woolley}{2022}]%
        {Woolley2022DigitalInfluencers}
\bibfield{author}{\bibinfo{person}{Samuel~C Woolley}.}
  \bibinfo{year}{2022}\natexlab{}.
\newblock \showarticletitle{{Digital Propaganda: The Power of Influencers}}.
\newblock \bibinfo{journal}{\emph{Journal of Democracy}} \bibinfo{volume}{33},
  \bibinfo{number}{3} (\bibinfo{year}{2022}), \bibinfo{pages}{115--129}.
\newblock
\urldef\tempurl%
\url{https://muse.jhu.edu/article/860232}
\showURL{%
\tempurl}


\bibitem[\protect\citeauthoryear{Wright, Shaikh, Park, Epperson, Ahmed, Pinel,
  Chau, and Yang}{Wright et~al\mbox{.}}{2021}]%
        {Wright2021Recast:Visualization}
\bibfield{author}{\bibinfo{person}{Austin~P. Wright}, \bibinfo{person}{Omar
  Shaikh}, \bibinfo{person}{Haekyu Park}, \bibinfo{person}{Will Epperson},
  \bibinfo{person}{Muhammed Ahmed}, \bibinfo{person}{Stephane Pinel},
  \bibinfo{person}{Duen Horng~(polo) Chau}, {and} \bibinfo{person}{Diyi Yang}.}
  \bibinfo{year}{2021}\natexlab{}.
\newblock \showarticletitle{{Recast: Enabling User Recourse and
  Interpretability of Toxicity Detection Models with Interactive
  Visualization}}.
\newblock \bibinfo{journal}{\emph{Proceedings of the ACM on Human-Computer
  Interaction}} \bibinfo{volume}{5}, \bibinfo{number}{CSCW1} (\bibinfo{date}{4}
  \bibinfo{year}{2021}).
\newblock
\showISSN{25730142}
\urldef\tempurl%
\url{https://doi.org/10.1145/3449280}
\showDOI{\tempurl}


\bibitem[\protect\citeauthoryear{Xia, Huang, Duan, and Whinston}{Xia
  et~al\mbox{.}}{2009}]%
        {Xia2009BallotCommunities}
\bibfield{author}{\bibinfo{person}{Mu Xia}, \bibinfo{person}{Yun Huang},
  \bibinfo{person}{Wenjing Duan}, {and} \bibinfo{person}{Andrew~B. Whinston}.}
  \bibinfo{year}{2009}\natexlab{}.
\newblock \showarticletitle{{Ballot box communication in online communities}}.
\newblock \bibinfo{journal}{\emph{Commun. ACM}} \bibinfo{volume}{52},
  \bibinfo{number}{9} (\bibinfo{date}{9} \bibinfo{year}{2009}),
  \bibinfo{pages}{138--142}.
\newblock
\showISSN{0001-0782}
\urldef\tempurl%
\url{https://doi.org/10.1145/1562164.1562199}
\showDOI{\tempurl}


\bibitem[\protect\citeauthoryear{Zannettou, Elsherief, Belding, Nilizadeh, and
  Stringhini}{Zannettou et~al\mbox{.}}{2020}]%
        {Zannettou2020MeasuringWebsites}
\bibfield{author}{\bibinfo{person}{Savvas Zannettou}, \bibinfo{person}{Mai
  Elsherief}, \bibinfo{person}{Elizabeth Belding}, \bibinfo{person}{Shirin
  Nilizadeh}, {and} \bibinfo{person}{Gianluca Stringhini}.}
  \bibinfo{year}{2020}\natexlab{}.
\newblock \showarticletitle{{Measuring and Characterizing Hate Speech on News
  Websites}}. In \bibinfo{booktitle}{\emph{12th ACM Conference on Web
  Science}}. \bibinfo{publisher}{ACM}, \bibinfo{address}{New York, NY, USA},
  \bibinfo{pages}{125--134}.
\newblock
\showISBNx{9781450379892}
\urldef\tempurl%
\url{https://doi.org/10.1145/3394231.3397902}
\showDOI{\tempurl}


\bibitem[\protect\citeauthoryear{Zhou and Farzan}{Zhou and Farzan}{2021}]%
        {Zhou2021DesigningCyberbullying}
\bibfield{author}{\bibinfo{person}{Yingfan Zhou} {and} \bibinfo{person}{Rosta
  Farzan}.} \bibinfo{year}{2021}\natexlab{}.
\newblock \showarticletitle{{Designing to Stop Live Streaming Cyberbullying}}.
  In \bibinfo{booktitle}{\emph{Proceedings of the 10th International Conference
  on Communities {\&} Technologies - Wicked Problems in the Age of Tech}}.
  \bibinfo{publisher}{ACM}, \bibinfo{address}{New York, NY, USA},
  \bibinfo{pages}{138--150}.
\newblock
\showISBNx{9781450390569}
\urldef\tempurl%
\url{https://doi.org/10.1145/3461564.3461574}
\showDOI{\tempurl}


\bibitem[\protect\citeauthoryear{Zuckerman}{Zuckerman}{2021}]%
        {Zuckerman2021WhyEcosystems}
\bibfield{author}{\bibinfo{person}{Ethan Zuckerman}.}
  \bibinfo{year}{2021}\natexlab{}.
\newblock \showarticletitle{{Why study media ecosystems?}}
\newblock \bibinfo{journal}{\emph{Information Communication and Society}}
  \bibinfo{volume}{24}, \bibinfo{number}{10} (\bibinfo{year}{2021}),
  \bibinfo{pages}{1495--1513}.
\newblock
\showISSN{14684462}
\urldef\tempurl%
\url{https://doi.org/10.1080/1369118X.2021.1942513}
\showDOI{\tempurl}


\end{thebibliography}

\appendix
\section{Codebook}
\label{codebook}
\begin{itemize}
    \item (0) Not relevant
    \item (1) Twitch sues hate raiders
    \item (2) Follow bots
    \item (3) Solutions to hate raid
    \item (4) Use HR movement to self promote
    \item (5) DayOffTwitch useless
    \item (6) Hate raider community outside Twitch
    \item (7) Recommended tools to combat hate raid
    \item (8) Suggestions to Twitch to combat hate raid
    \item (9) Ineffective tools 
    \item (10) Official Twitch tools to combat hate raid
    \item (11) Follow bots grab IPs
    \item (12) Definition of hate raid
    \item (13) Small streamers targeted
    \item (14) Call for cultural change among Twitch users
    \item (15) Streamers share stories
    \item (16) Speculation Twitch users leave to join competitors
    \item (17) Ghost viewers
    \item (18) People expecting too much from Twitch
    \item (19) Minority streamers being targeted
    \item (20) Psychological impact of hate raids
    \item (21) Engagement impact of hate raids
    \item (22) Relevant but not in list
\end{itemize}

 \section{Tools Descriptions with Users' Opinions}
\label{tools}
\begin{table}[h]
  \caption{Tools with Description and User's Attitudes}
  \label{tab:commands}
  \scalebox{.75}{\begin{tabular}{p{2.5cm}p{4.5cm}p{4.5cm}p{4.5cm}}
    \toprule
    Tool/Feature &Description & What Supporters Believe & What Opponents Believe\\
    \midrule
     Automod & Automod is provided by Twitch to prevent harmful chat messages from being seen by other viewers and includes levels of moderation and a customizable word block list.    &
     \begin{itemize}[leftmargin=*]
        \item Built in tool - easy to set up 
         \item Automatically catch harmful messages
         \item Moderator has to manually ban/mute, so mistakes are easily resolved 
    \end{itemize}
      
 & 
 \begin{itemize}[leftmargin=*]
     \item Can't automatically ban/mute, so streamer has to be actively involved
     \item Word filters can be easily circumvented
 \end{itemize}

\\ 
Twitch Chat Settings & Twitch chat settings modify who is allowed to send messages in chat and what those messages are like - for example, a streamer or moderator can set their chat to only accept messages from those who are followers. 
& \begin{itemize}[leftmargin=*]
    \item Built in features - easy to set up 
    \item Can prevent attackers from sending harmful messages 
\end{itemize}
& \begin{itemize}[leftmargin=*]
    \item Has to be manually modified for different situations
    \item Safer (more restrictive) chat settings can decrease engagement 
\end{itemize}
    \\
Verification & Twitch allows streamers to  allow only those who have verified their accounts (by email, phone, or both) to send messages.
    &
    \begin{itemize}[leftmargin=*]
        \item Attackers have to spend more time verifying bot accounts
        \item If one account linked to an email is banned from a channel, all accounts linked to that email are as well
    \end{itemize}
     &
     \begin{itemize}[leftmargin=*]
         \item It is easy to create many emails to verify bots
         \item Possible (though more difficult) to do same with phone numbers
         \item Most importantly, these decrease engagement
     \end{itemize}

 \\ Reviewing Extensions & Twitch allows channels to use extensions to enhance their stream (such as captions). Some extensions contact external servers , which could allow an attacker to log a viewer's IP address. Because of this, some users wanted Twitch to manually review every extension for security issues. 
 & \begin{itemize}[leftmargin=*]
     \item Reviewing these extensions would reduce the chance of personal information being leaked
 \end{itemize}
 
 & \begin{itemize}[leftmargin=*]
     \item Extensions are already reviewed before they are released
     \item Everything you connect to can see your IP, and it isn't very dangerous, so this is a waste of resources
    \end{itemize} 
\\

\\
IP bans & While banning individual bot accounts is pointless, banning IP addresses associated with a set of bot accounts would take them all offline at once. 

& \begin{itemize}[leftmargin=*]
    \item More efficient and less frustrating than banning individual bots
    \item Twitch already sometimes IP bans reported accounts
\end{itemize}
& \begin{itemize}[leftmargin=*]
    \item IPs are dynamic and change, so the bot could one day receive a new, unbanned IP
    \item If these accounts use VPNs, they can easily change their IP address 
\end{itemize}

\\ 

Increasing Streamer Control Over Raids & Raiding is an important feature on Twitch, even if it is sometimes misused for hate raids (though most hate raids do not actually utilize the raid feature). Streamers can either reject all raids, allow only from friends, or allow from everyone. 
& \begin{itemize}[leftmargin=*]
    \item Restricting or disabling raids hinders channel growth
    \item Accepting all raids leaves streamers vulnerable
    \item Allowing streamers to accept or reject individual raids would avoid both of these issues
\end{itemize}
& \begin{itemize}[leftmargin=*]
    \item This will have a small impact on hate raids
\end{itemize}

\\ Third Party Tools & Third party tools are bots created to help streamers and moderators moderate their streams and provide other services unrelated to moderation. These are not made by Twitch, but the most popular ones have hundreds of thousands of users. 
& \begin{itemize}[leftmargin=*]
    \item Gives streamers more flexibility and control
    \item Gives streamers peace of mind 
\end{itemize}

& \begin{itemize}[leftmargin=*]
    \item Many commenters were frustrated that third parties were working on problems that they believed Twitch was ignoring
\end{itemize}

\\
 \bottomrule
  \end{tabular}}
\end{table}

\end{document}